\documentclass[12pt,a4paper]{article}

\usepackage{amsmath,amssymb} 
\usepackage{accents}
\usepackage{color}
\usepackage[bookmarks=true,hyperfigures=true]{hyperref}
\usepackage{graphicx}
\usepackage[nosort]{cite}
\usepackage[bulletsep]{collref}
\usepackage{tensor}
\usepackage{shuffle}
\usepackage{lscape}
\usepackage{empheq}

\usepackage{color}

\usepackage[a4paper,text={173mm,216mm},centering]{geometry}

\allowdisplaybreaks[3]

\usepackage[font=small,labelfont=bf,width=0.85\textwidth]{caption}

\let\oldbfseries=\bfseries
\let\oldmdseries=\mdseries
\let\oldnormalfont=\normalfont
\renewcommand{\bfseries}{\oldbfseries\boldmath}
\renewcommand{\mdseries}{\oldmdseries\unboldmath}
\renewcommand{\normalfont}{\oldnormalfont\unboldmath}

\makeatletter
\newlength{\apb@width}
\newcommand{\autoparbox}[2][c]{\settowidth{\apb@width}{#2}\parbox[#1]{\apb@width}{#2}}

\makeatother

\newcommand{\nn}{\nonumber}

\newcommand{\remark}[2][.]{{\color{red}\renewcommand{\bfdefault}{b}\rmfamily\if.#1\else\textbf{#1:} \fi#2}}

\newcommand{\be}{\begin{equation}}
\newcommand{\ee}{\end{equation}}
\newcommand{\beq}{\begin{equation}}
\newcommand{\eeq}{\end{equation}}

\newcommand{\bma}{\begin{pmatrix}}
\newcommand{\ema}{\end{pmatrix}}
\newcommand{\ba}{\begin{eqnarray}}
\newcommand{\ea}{\end{eqnarray}}
\newcommand{\trans}{{\scriptscriptstyle\mathrm{T}}}         
 
\newcommand{\A}{\mathcal{A}}   

\ifx\genfrac\sdflkaj\else\fi

\newcommand{\cA}{\mathcal{A}}

\newcommand{\cC}{\mathcal{C}}
\newcommand{\cP}{\mathcal{P}}

\newcommand{\mcC}{\mathcal{C}}

\newcommand{\ep}{\epsilon}

\def\l<{\langle}\def\r>{\rangle}

\makeatletter
\newcommand{\namedref}[2]{\hyperref[#2]{#1~\ref*{#2}}}
\newcommand{\secref}{\@ifstar{\namedref{Section}}{\namedref{sec.}}}
\newcommand{\subsecref}{\@ifstar{\namedref{Subsection}}{\namedref{subsec.}}}
\newcommand{\appref}{\@ifstar{\namedref{Appendix}}{\namedref{app.}}}
\newcommand{\tabref}{\@ifstar{\namedref{Table}}{\namedref{tab.}}}
\newcommand{\figref}{\@ifstar{\namedref{Figure}}{\namedref{fig.}}}
\makeatother

\newif\ifmrnote 
\mrnotetrue
 
\usepackage[active]{srcltx}

\newif\ifjbnote 
\jbnotetrue
 
\newcommand{\matr}[2]{\left(\begin{array}{#1}#2\end{array}\right)}

\def\[{\begin{equation}}
\def\]{\end{equation}}
\def\<{\begin{eqnarray}}
\def\>{\end{eqnarray}}

\newcommand{\Tr}{\mathop{\mathrm{Tr}}}

\newcommand{\eqn}[1]{(\ref{#1})}

\newcommand\insertleg[3]{\frac{\sigma_{#1,#3}}{\sigma_{#1,#2}\,\sigma_{#2,#3}}}
\newcommand\inserttleg[4]{\frac{\sigma_{#1,#4}}{\sigma_{#1,#2}\,\sigma_{#2,#3}\,\sigma_{#3,#4}}}

\def\bea{\begin{align}}
\def\eea{\end{align}}
\def\be{\begin{equation}}
\def\ee{\end{equation}}

\begin{document}
\thispagestyle{empty}

\begingroup\raggedleft\footnotesize\ttfamily
HU-EP-16/22\\
\vspace{15mm}
\endgroup

\begin{center}
{\LARGE\bfseries Einstein-Yang-Mills from  \\pure Yang-Mills amplitudes
\par}%

\vspace{15mm}

\begingroup\scshape\large 
Dhritiman Nandan${}^{1}$, Jan Plefka${}^{1}$, Oliver Schlotterer${}^{2}$, \\ Congkao Wen${}^{3}$
\endgroup
\vspace{5mm}

\textit{${}^{1}$ Institut f\"ur Physik and IRIS Adlershof, Humboldt-Universit\"at zu Berlin, \phantom{$^\S$}\\
  Zum Gro{\ss}en Windkanal 6, D-12489 Berlin, Germany} \\[0.1cm]
  \textit{${}^{2}$ 
  Max-Planck-Institut f\"ur Gravitationsphysik, Albert-Einstein-Institut \\ 
  Am M\"uhlenberg 1, D-14476 Potsdam, Germany
  } \\[0.1cm]
\textit{${}^{3}$ I.N.F.N. Sezione di Roma Tor Vergata, Via della Ricerca Scientifica\\
 00133 Roma, Italy} \\[0.1cm]

\bigskip
  
\texttt{\small\{dhritiman.nandan, jan.plefka\}@physik.hu-berlin.de,} 

\texttt{olivers@aei.mpg.de,  congkao.wen@roma2.infn.it} 
\vspace{8mm}


\textbf{Abstract}\vspace{5mm}\par
\begin{minipage}{14.7cm}

We present new relations for scattering amplitudes of color ordered gluons and gravitons in Einstein-Yang-Mills theory. 
Tree-level amplitudes of arbitrary multiplicities and polarizations involving up to three gravitons and up to two color traces are 
reduced to partial amplitudes of pure Yang-Mills theory. In fact, the double-trace identities apply to Einstein-Yang-Mills extended 
by a dilaton and a B-field. Our results generalize recent work of Stieberger and Taylor for the single graviton case with a single 
color trace. As the derivation is made in the dimension-agnostic Cachazo-He-Yuan formalism, our results are valid for external 
bosons in any number of spacetime dimensions. Moreover, they generalize to the superamplitudes in theories with 16 
supercharges. \end{minipage}\par

\end{center}
\newpage



\setcounter{tocdepth}{1}


\tableofcontents

\section{Introduction}

Einstein's theory of gravity and Yang-Mills (YM) gauge theories are both built
on local symmetries yet their dynamical structure is quite different. Nonetheless,
in a perturbative quantization of these theories in a flat space-time background
intimate relations between their S-matrices have been 
uncovered that are far from obvious at the Lagrangian level. They allow to 
express graviton scattering amplitudes through YM scattering data, being
entirely unobvious from a Feynman-diagram based computation. 
The first such connection are the Kawai-Lewellen-Tye (KLT) relations \cite{Kawai:1985xq} 
derived from the string theoretic origin of the tree-level field-theory S-matrices. 
The KLT relations express graviton amplitudes as sums of products of two color ordered gluon amplitudes. 

More recently, Bern, Carrasco and Johansson (BCJ) \cite{Bern:2008qj, Bern:2010yg, Bern:2010ue}
introduced a double-copy construction of graviton amplitudes through gluon amplitudes of the
same multiplicity.
Here Lie-algebra like relations for the kinematic building blocks 
of gauge-theory amplitudes were identified and used to construct graviton amplitudes.
This technique has proven to be enormously powerful to generate loop-level integrands
of gravitational theories from the simpler gauge-theory ones and became the state-of-the-art method
to explore the UV properties of supergravities, see for instance \cite{Bern:2012uf, Bern:2012cd, Bern:2013uka, Bern:2014sna}. 
At tree level the BCJ double-copy construction enforces the so-called BCJ relations between 
color ordered gluon tree
amplitudes reducing the basis of independent $n$-gluon amplitudes to $(n-3)!$ entries \cite{Bern:2008qj}.
These BCJ relations have been proven in a variety of different ways, including monodromy properties of the string worldsheet \cite{BjerrumBohr:2009rd,Stieberger:2009hq}, the field-theory limit of open-superstring tree amplitudes \cite{Mafra:2011nv, Mafra:2011nw}, BCFW on-shell recursions \cite{Feng:2010my, Chen:2010ct, Chen:2011jxa} and cohomology arguments in pure-spinor superspace \cite{Mafra:2015vca}. 

Less is known about the explicit S-matrix elements for mixed graviton and gluon scattering in 
Einstein gravity minimally coupled to YM theory, EYM for short. In the 1990s gravitationally 
dressed amplitudes in four dimensions for the maximally-helicity violating (MHV) case were given
in \cite{Selivanov:1997aq,Selivanov:1997ts,Bern:1999bx}, where at most two gluons or
one gluon and one graviton have opposite helicities to the other particles. These results
were established using a self-dual classical `perturbiner' solution \cite{Selivanov:1997aq,Selivanov:1997ts}
or by again employing the KLT relations \cite{Bern:1999bx}. Double-copy constructions for gluon-graviton
scattering in supergravity theories were given in \cite{Chiodaroli:2014xia,Chiodaroli:2015rdg}, also see \cite{Chiodaroli:2016jqw} for a review.

Very recently a nice and compact formula relating the scattering of a single graviton with
$n$ color ordered gluons of arbitrary helicities to a linear combination of ($n+1)$-gluon 
scattering amplitudes was found by Stieberger and Taylor \cite{Stieberger:2016lng}. Their derivation is based on
a new set of monodromy relations for mixed open-closed string amplitudes along the lines of
\cite{Stieberger:2009hq, Stieberger:2015vya}. With $p^{\mu}$ denoting the graviton momentum, their formula reads
\be
\A_{\text{EYM}}(1,2,\ldots,n;p)= \frac{\kappa}{g}\, \sum_{l=1}^{n-1}\epsilon_{p}\cdot x_{l}\,
\A(1,2,\ldots, l,p,l+1,\ldots,n)\, ,
\label{master}
\ee
where $\kappa$ and $g$ are the gravitational and YM couplings, respectively. 
Note that this amplitude is associated to a single trace color structure of the form
$\Tr( T^{a_{1}}T^{a_{2}}\ldots T^{a_{n}})$, where $T^a$ denote the generators of the non-abelian gauge group.
Moreover, $x_{l}$ denotes the region momentum 
\be
x_{l}^{\mu}\equiv \sum_{j=1}^{l}k_{l}^{\mu}
\ee
of the gluons with lightlike momenta $k^{\mu}_{l}$.
In this work we shall give a concise field-theory proof of the relation \eqn{master} 
and extend it to more (up to three) graviton insertions and higher
(up to two) color trace structures.

In four spacetime dimensions,
the development of on-shell techniques\footnote{See \cite{Dixon:1996wi,Henn:2014yza,Elvang:2013cua} for introductory references.} has given us powerful representations of tree-level gluon and also graviton
amplitudes at hand. In fact there are now closed analytical expressions available for all
color ordered $n$-gluon amplitudes \cite{Drummond:2008cr} as well as graviton amplitudes \cite{Drummond:2009ge} at tree level based on solving the BCFW recursion \cite{Britto:2005fq} in its supersymmetric extension \cite{Brandhuber:2008pf,ArkaniHamed:2008gz,Elvang:2008na} which have been implemented in computer algebra packages \cite{Dixon:2010ik,Bourjaily:2010wh,Schuster:2013aya}. 
On-shell recursion relations have been further extended for planar loop integrands \cite{ArkaniHamed:2010kv}. 

In higher spacetime dimensions, recent progress has been driven by different sets of methods:
The Berends-Giele recursion \cite{Berends:1987me} for an efficient resummation of Feynman diagrams 
and the field-theory limit of string amplitudes, starting with \cite{Green:1982sw}. 
The pure-spinor formalism of the superstring \cite{Berkovits:2000fe} 
inspired a recursive setup to determine manifestly supersymmetric $n$-gluon tree amplitudes in ten-dimensional super Yang-Mills (SYM) from supersymmetry, gauge invariance and locality \cite{Mafra:2010jq}. An extension of these methods to loop level has been initiated in \cite{Mafra:2014gja, Mafra:2015mja}. Machine-readable component expressions in ten dimensions are significantly facilitated by the techniques in \cite{Lee:2015upy} and available for download on \cite{PSwebsite}. Moreover, the pure-spinor approach allows for an alternative proof of BCJ relations \cite{Mafra:2015vca} as well as explicit constructions of BCJ numerators at tree level \cite{Mafra:2011kj} and loop level \cite{Mafra:2014gja, Mafra:2015mja}.

Another line of attack towards the higher-dimensional tree level S-matrix for gravitons, gluons, cubic scalars and beyond is provided by the Cachazo-He-Yuan (CHY) formalism \cite{Cachazo:2013gna, Cachazo:2013hca, Cachazo:2013iea, Cachazo:2014nsa, Cachazo:2014xea}. As suggested by their origin from ambitwistor strings \cite{Mason:2013sva, Casali:2015vta} and related recent developments \cite{Siegel:2015axg, Huang:2016bdd, Casali:2016atr}, CHY formulae yield unifying representations for a variety of tree amplitudes\footnote{See \cite{Adamo:2013tsa, Casali:2014hfa, Adamo:2015hoa, Geyer:2015bja, He:2015yua, Geyer:2015jch, Cardona:2016bpi, Cardona:2016wcr} for extensions to loop level.} which strongly resemble those of the superstring. For instance, the pure-spinor incarnation of the CHY formalism \cite{Berkovits:2013xba} is know from \cite{Gomez:2013wza} to reproduce the supersymmetric tree amplitudes from the field-theory limit of the superstring \cite{Mafra:2010jq, Mafra:2011nv}.

Similar to string theory, CHY formulae compactly represent amplitudes in Einstein gravity, pure YM and cubic massless scalar theories in arbitrary dimensions in terms of an integral over the punctured sphere. These CHY integrals localize on the solutions to the scattering equations 
\be
f_a \equiv \sum_{b=1 \atop b\not=a}^{n} \frac{s_{ab}}{\sigma_a - \sigma_b} =0  \; ,
\qquad \text{where} \quad s_{ab}\equiv k_a \cdot k_b\, ,
\label{eq-scateq}
\ee
where $k_{a}^{\mu}$ denote the light-like momenta and the $\sigma_{a}$ the positions
of the punctures. Such CHY integrals yield the same propagators as seen in the field-theory limit 
of worldsheet integrals in string theory \cite{Broedel:2013tta, Cachazo:2013iea}, see \cite{Mafra:2016ltu} for 
an efficient recursion via Berends-Giele currents.

In this work, we will employ the CHY formalism, in particular the results of \cite{Cachazo:2014nsa, Cachazo:2014xea}, to derive the EYM relation \eqn{master}
and its generalizations. Our results therefore hold in arbitrary spacetime dimensions and by pure-spinor methods \cite{Mafra:2010jq, Mafra:2011nv, Berkovits:2013xba, Gomez:2013wza} extend to any superamplitude descending from ten-dimensional SYM coupled to half-maximal supergravity. The key idea is to rewrite the graviton building blocks in the CHY integrand
in terms of so-called Parke-Taylor factors, thus reducing the graviton-gluon amplitudes
to linear combinations of polarization-dependent sums of gluon amplitudes. An almost identical derivation
can be performed in the heterotic string which is left for future work.

Our paper is organized as follows. In section \ref{section:CHY} we give a brief review of the 
CHY representations of tree amplitudes in general and focus on the integrand for mixed gluon-graviton 
scattering. Section \ref{section:1graviton} proves \eqn{master} whereas section \ref{sec4} and \ref{sec5} generalize \eqn{master}
to the two- and three-graviton case. The EYM amplitude relations we find include\footnote{Note that the $i=1$ contributions in the second and third line of (\ref{twogravintro}) are understood as ${\cal A}(q,1,2,\ldots,n)$. Moreover, subamplitudes with adjacent gravitons occur in each line of (\ref{twogravintro}) including the terms
$\sum_{i=1}^{n-1}(\ep_p\cdot x_i) $
$(\ep_q\cdot x_i)[ {\cal A}(1,\ldots,i,p,q,i+1,\ldots,n)+{\cal A}(1,\ldots,i,q,p,i+1,\ldots,n)]$
from the double sum in the first line.} 
\begin{align} 
\cA_{\text{EYM}} (1,2, \ldots,n; p,q)\
&= \frac{\kappa^{2}}{g^{2}}\Bigl [
\sum_{1=i\leq j}^{n-1}   (\epsilon_{p}\cdot x_{i})\, (\epsilon_{q}\cdot x_{j})\, 
\cA(1,\ldots,i,p,i{+}1,\ldots, j, q,j{+}1,\ldots, n)  \notag \\
&\! \! \!  \! \! \!  \! \! \!  \! \! \!  \! \! \! \! \! \! \! \! \! \! \! \! \! 
 - (\epsilon_{q} \cdot p)  \sum_{j=1}^{n-1}
  (\epsilon_{p}  \cdot x_{j}) \sum_{i=1}^{j+1} 
\,
\cA(1,2,\ldots,i{-}1,q, i, \ldots, j,p, j{+}1,\ldots,n) 
 \label{twogravintro} \\
&  \! \! \!  \! \! \!  \! \! \!  \! \! \!  \! \! \! \! \! \! \! \! \! \! \! \! \! 
- { (\epsilon_{p}  \cdot  \epsilon_{q}) \over 2} \sum_{l=1}^{n-1} 
(p\cdot k_l) \sum_{1=i\leq j}^l 
\cA( 1,2,\ldots, i{-}1,q,i, \ldots,j{-}1,p,j,\dots ,n) 
+ (  p \leftrightarrow q)\, \Bigr ] \, , \notag 
\end{align}
for $n$ gluons with momenta $k_l$ and two gravitons with momenta $p,q$. The analogous
three-graviton identity is given in (\ref{threegr}). Moreover, the $\mathcal{A}(\ldots)$ and the
$\mathcal{A}_{\text{EYM}}(\ldots)$ may be read as
superamplitudes in ten-dimensional SYM and supersymmetrized EYM theories with 16 supercharges, respectively.
In section \ref{sec6} we comment on the four- and higher-graviton
cases indicating that there are no conceptual problems to resolve them. In section \ref{section:multitrace} and 
\ref{section:moremultitrace} we turn to the multi-trace amplitudes in EYM augmented by a B-field and a dilaton
with the main results in (\ref{mtrace2}) and (\ref{doubletraceonegravgeneral})
before ending with an outlook.

\section{CHY representation of scattering amplitudes} \label{section:CHY}

In terms of the CHY formula \cite{Cachazo:2013gna, Cachazo:2013hca, Cachazo:2013iea, Cachazo:2014nsa, Cachazo:2014xea} the scattering amplitude for $n$ massless particles with momenta $k_{a}$ and 
polarizations $\epsilon_{a}$ takes the general form
\be
\label{CHYmaster}
  \mathcal{A}_{n} = \int d\mu_{n}\prod_{a=1}^{n}\!{}'\, \delta(f_a) \, 
  \mathcal{I}_{n}(\{k,\epsilon,\sigma\}) \; .
\ee
The integration with measure $ d\mu_{n}\equiv \frac{ d^{n}\sigma_a }{\text{vol SL(2,$\mathbb{C}$)}} $ is performed over 
the moduli space of punctured spheres, and the
$\delta$-functions enforcing the scattering equations \eqn{eq-scateq} completely localize the integrals.
Here one needs to divide by the volume of SL(2,$\mathbb{C})$ as the integrand is invariant under M\"obius transformations. 
In the following we abbreviate the measure as
\be
d\Omega_{n}\equiv d\mu_{n}\prod_{a=1}^{n}\!{}'\, \delta(f_a) \,.
\label{measOMEGA}
\ee
Moreover, in the expression above $\prod'$
refers to the fact that one needs to remove three delta functions in a way explained in \cite{Cachazo:2013gna, Cachazo:2013hca, Cachazo:2013iea, Cachazo:2014nsa,Cachazo:2014xea}. The integrand for a specific bosonic theory,  $\mathcal{I}_{n}(\{k,\epsilon,\sigma\})$, is constructed from a combination of building blocks. We set the couplings $\kappa$ and $g$ to unity from now on.
These are on the one hand the Park-Taylor factor  
\be
\cC(1,\ldots, n)\equiv \frac{1}{\sigma_{1,2}\sigma_{2,3}\ldots \sigma_{n,1}}\, , \qquad \sigma_{a,b} \equiv \sigma_{a}-\sigma_{b}
\ee
and on the other hand the reduced Pfaffian of an anti-symmetric $2n\times 2n$
matrix $\Psi$, 
\be
\text{Pf}'\, \Psi_{n}(\{k,\epsilon,\sigma\})\, , \qquad \text{where}\quad
  \Psi_{n} \equiv \matr{cc}{A & -C^\trans \\ C & B}
  \label{prime}
\ee
with the entries
\be
\label{ABCmatrices}
  A_{ab} \equiv \begin{cases}
    \frac{k_a\cdot k_b}{\sigma_a - \sigma_b} & a\not=b \; , \\
    0                                        & a=b\, , 
  \end{cases}
  \qquad
  B_{ab} \equiv \begin{cases}
    \frac{\epsilon_a\cdot \epsilon_b}{\sigma_a - \sigma_b} & a\not=b \; , \\
    0                                        & a=b\, , 
  \end{cases}
  \qquad
  C_{ab} \equiv \begin{cases}
    \frac{\epsilon_a\cdot k_b}{\sigma_a - \sigma_b} & a\not=b \; , \\
    - \sum\limits_{c\not=a} \frac{\epsilon_a\cdot k_c}{\sigma_a - \sigma_c} & a=b\, . 
  \end{cases}
\ee
The prime along with the Pfaffian in (\ref{prime}) instructs to remove any two rows and columns $i,j$, i.e.
\beq
\text{Pf}'\, \Psi_{n}(\{k,\epsilon,\sigma\}) \equiv \frac{  (-1)^{i+j} }{\sigma_i-\sigma_j} \text{Pf} \big[ \Psi_{n}(\{k,\epsilon,\sigma\}) \big]^{ij}_{ij} \ ,
\label{reducePF}
\eeq
where $[\ldots ]^{ij}_{ij}$ is obtained from the enclosed matrix by deleting the $i^{\rm th}$ and $j^{\rm th}$ row and 
column, respectively. We note that, on the support of the scattering equations, the Pfaffian is invariant under permutations 
of any of the $n$ particles and independent on this choice of $i$ and $j$. Combining these two building blocks in the 
integrand, one may write a pure gluon amplitude using
\begin{align}
\mathcal{I}_{n}^{\text{YM}}(1,2,\ldots,n)&= \cC(1,2,\ldots, n)\, \text{Pf}'\, \Psi_{n}(\{k,\epsilon,\sigma\})  \ ,
\label{pleaseSUSY}
\end{align}
whereas the single-trace part of an EYM amplitude with a single graviton at position $n+1$ 
with momentum $p$ takes an integrand of the form \cite{Cachazo:2014nsa}
\be
\mathcal{I}_{n+1}^{\text{EYM}}(1,2,\ldots,n;p)=  \cC(1,2,\ldots, n)\, C_{pp} \,
\text{Pf}'\,  \Psi_{n+1}( \{k_{a},p,\epsilon,\sigma\}) \, .
\ee
Using these expressions it is straightforward to prove \eqn{master} and extend it to more
graviton insertions. 

The single-trace sector of a general $r$-graviton and $n$-gluon EYM amplitude has the compact CHY representation \cite{Cachazo:2014nsa}
\be
\mathcal{I}_{n+r}^{\text{EYM}}(1,2,\ldots,n;p_{1},\ldots, p_{r})=  
 \cC(1,2,\ldots, n)\, \text{Pf} \, \Psi_{r}( \{p,\epsilon_{p},\sigma\})
  \text{Pf}'\,  \Psi_{n+r}( \{k,p,\epsilon_{k},\epsilon_{p},\sigma\}) \, ,
\label{EYM-CHY-gen}
\ee
where $\Psi_{r}(\{p,\epsilon_{p},\sigma\})$ is a CHY matrix extending only over the $r$
graviton legs without any deletions as in (\ref{reducePF}). Double-trace generalizations of (\ref{EYM-CHY-gen}) can be found
in sections \ref{section:multitrace} and \ref{section:moremultitrace}. In order to rewrite EYM amplitudes in terms of color ordered gluon amplitudes with integrands (\ref{pleaseSUSY}) one simply needs to seek identities of the
schematic kind
\be
 \cC(1,2,\ldots, n)\,  \text{Pf} \, \Psi_{r}(\{p,\epsilon_{p},\sigma\})
=\sum_{\cP_{i}\in\text{Perm}(1,\ldots,n,p_{1},\ldots,p_{r})}
F_{\cP_{i}}(\epsilon_{p},p,k)\, \cC(\cP_{i})\, ,
\label{desired}
\ee
which we shall provide with explicitly known functions $F_{\cP_{i}}$ of the polarizations and momenta.

\section{One graviton} \label{section:1graviton}

In order to derive the single-graviton relation
\eqn{master} from the CHY formalism, we start from its right hand side and note that, 
by the permutation invariance of $\text{Pf}'\, \Psi_{n+1}$ under the measure $d\Omega_{n+1}$ in (\ref{measOMEGA}),
\be
\A(1,2,\ldots, l,p,l{+}1,\ldots,n) = \int d\Omega_{n+1}\, 
\frac{\sigma_{l,l+1}}{\sigma_{l,p}\, \sigma_{p,l+1}}\, 
\frac{1}{\sigma_{1,2}\ldots \sigma_{l,l+1} \ldots \sigma_{n,1}}\, \text{Pf}'\,  \Psi_{n+1}(\{k,p,\epsilon,\sigma\})  \, .
\ee
Hence, given the EYM integrand (\ref{EYM-CHY-gen}), all there is to do in order to prove \eqn{master} is to show that
\be
\text{Pf} \, \Psi_{r=1}= C_{pp} =\sum_{l=1}^{n-1} \epsilon_{p}\cdot x_{l}\, \frac{\sigma_{l,l+1}}{\sigma_{l,p}\, \sigma_{p,l+1}}
\, ,
\label{Cqq-1g}
\ee
which is elementary. Writing $\frac{\sigma_{l,l+1}}{\sigma_{l,p}\, \sigma_{p,l+1}}=
\frac{1}{\sigma_{l,p}}-\frac{1}{\sigma_{l+1,p}}$ we have a telescoping sum
\be
C_{pp}= \sum_{l=1}^{n-1}\, \Bigl ( 
\frac{\epsilon_{p}\cdot x_{l}}{\sigma_{l,p} }- \frac{\epsilon_{p}\cdot x_{l}}{\sigma_{l+1,p}}
\Bigr ) = 
\sum_{l=2}^{n-1} \frac{\epsilon_{p}\cdot (x_{l}-x_{l-1})}{\sigma_{l,p}}
+ \frac{\epsilon_{p}\cdot k_{1}}{\sigma_{1,p}} - \frac{\epsilon_{p}\cdot x_{n-1}}{\sigma_{n,p}}\, .
\ee
Now using $x_{l}-x_{l-1}=k_{l}$ and $-x_{n-1}=k_{n}{+}p$ from total momentum
conservation, along with $\epsilon_{p}\cdot p=0$, we indeed reproduce the diagonal element of the $C$-matrix in (\ref{ABCmatrices})
\be
C_{pp}=\sum_{l=1}^{n} \frac{\epsilon_{p}\cdot k_{l}}{\sigma_{l,p}}  \, ,
\ee
which proves \eqn{master}. 

\section{Two gravitons} 
\label{sec4}

In order to address the two-graviton problem we first note that the relevant
Pfaffian of $\Psi_{r}$ in \eqn{EYM-CHY-gen} takes the form (writing $p_{1}\equiv p$ and $p_{2}\equiv q$)
\be
\text{Pf} \,\Psi_{r=2}= C_{pp}\, C_{qq}\, - \frac{s_{pq}\, 
(\epsilon_{p}\cdot \epsilon_{q})}{\sigma_{p,q}^{2}} + \frac{(\epsilon_{p}\cdot q)
\,(\epsilon_{q}\cdot p)}{\sigma_{p,q}^{2}} \, , \qquad s_{pq} \equiv p\cdot q\, .
\label{2gmaster}
\ee
The inequivalent tensor structures for the graviton polarizations can be conveniently classified after
rearranging the first term in (\ref{2gmaster}) via %
\begin{align}
C_{pp}= \sum_{i=1}^{n-1}(\epsilon_{p}\cdot x_{i}) \frac{\sigma_{i,i+1}}{\sigma_{i,p}\,
\sigma_{p,i+1}} + (\epsilon_{p}\cdot q )\insertleg{q}{p}{n}\, , 
\label{Cpp}\\
C_{qq}= \sum_{i=1}^{n-1}(\epsilon_{q}\cdot x_{i}) \frac{\sigma_{i,i+1}}{\sigma_{i,q}\,
\sigma_{q,i+1}} + (\epsilon_{q}\cdot p) \insertleg{p}{q}{n}\, .
\label{Cqq}
\end{align}
This generalizes relation \eqn{Cqq-1g} to two gravitons and $n$
gluons. Multiplying these two terms as they enter \eqn{2gmaster}, we arrive at
\begin{align}
\text{Pf} \,\Psi_{r=2}= C'_{pp}\, C'_{qq} +
C'_{pp}\,
(\epsilon_{q}\cdot p) \insertleg{p}{q}{n}
+ C'_{qq}\,
(\epsilon_{p}\cdot q )\insertleg{q}{p}{n}
 \, - \frac{s_{pq}\, 
(\epsilon_{p}\cdot \epsilon_{q})}{\sigma_{p,q}^{2}}
\label{2gmasteralt}
\end{align}
with
\be \label{eq:cprime}
C_{pp}' \equiv \sum_{i=1}^{n-1}(\epsilon_{p}\cdot x_{i}) \frac{\sigma_{i,i+1}}{\sigma_{i,p}\,
\sigma_{p,i+1}} \ .
\ee
Note that the tensor structure $(\epsilon_{q}\cdot p) \, (\epsilon_{p}\cdot q)$ cancels between 
$C_{pp}\, C_{qq}$ and the last term in \eqn{2gmaster}. In the following, we rearrange the three classes of terms in 
 (\ref{2gmasteralt}) such that a superposition of $(n+2)$-particle Parke-Taylor factors as in (\ref{desired}) arises upon multiplication with $\cC(1,2,\ldots, n)$. 
 
\subsection{Two-graviton contributions $(\epsilon_{p}\cdot x_{i}) (\epsilon_{q}\cdot x_{j})$}

The first term in (\ref{2gmasteralt}) is a product of two expressions as in (\ref{eq:cprime}),
\begin{align}
C_{pp}'\, C_{qq}' \, \cC(1,2,\ldots, n)
&=
\sum_{i,j=1\atop i\neq j}^{n-1} (\epsilon_{p}\cdot x_{i})\,(\epsilon_{q}\cdot x_{j})\,
\cC(1,2,\ldots,i,p,i+1,\ldots,j,q,j+1,\ldots,n) \notag \\ & \quad 
+
\sum_{i=1}^{n-1}(\epsilon_{p}\cdot x_{i})\,(\epsilon_{q}\cdot x_{i})
\, \frac{\sigma_{i,i+1}^{2}}{\sigma_{i,p}\, \sigma_{p,i+1}\, \sigma_{i,q}\,\sigma_{q,i+1}}
\, \cC(1,2,\ldots,n)\, .
\end{align}
While the first line is already in the desired form (\ref{desired}), the second line needs a small rearrangement.
Multiplying it with the identity $1=\sigma_{p,q}/\sigma_{p,q}$ and using the analogue
of Schouten's identity
\be
\sigma_{i,i+1}\, \sigma_{p,q} = - \sigma_{i,p}\, \sigma_{q,i+1} + \sigma_{i,q}\, \sigma_{p,i+1} \ ,
\ee
we straightforwardly establish the identity
\be
\frac{\sigma_{i,i+1}^{2}\, \sigma_{p,q}}{\sigma_{i,p}\, \sigma_{p,i+1}\, \sigma_{i,q}\,\sigma_{q,i+1}\, \sigma_{pq}}
= \inserttleg{i}{p}{q}{i+1} + \inserttleg{i}{q}{p}{i+1}\, .
\ee
Using this we thus have
\begin{align}
C_{pp}'\, C_{qq}' &\, \cC(1,2,\ldots, n)=
\sum_{i,j=1\atop i\neq j}^{n-1} (\epsilon_{p}\cdot x_{i})\,(\epsilon_{q}\cdot x_{j})\,
\cC(1,2,\ldots,i,p,i+1,\ldots,j,q,j+1,\ldots,n) \notag \\ &  
+
\sum_{i=1}^{n-1}(\epsilon_{p}\cdot x_{i})\,  (\epsilon_{q}\cdot x_{i})\Bigl [\, 
\cC(1,2,\ldots,i,p,q,i+1,\ldots,n) + ( p \leftrightarrow q) \, \Bigr ]\, ,
\end{align}
completing the rearrangement of this contribution into the desired form (\ref{desired}).

\subsection{Two-graviton contributions $(\epsilon_{p}\cdot x_{i}) (\epsilon_{q}\cdot p)$}
\label{sec42}

For the cross-terms arising in the product  $C_{pp}\, C_{qq}$ with (\ref{Cpp}) and (\ref{Cqq}), one has
\beq
C_{pp}'\, (\epsilon_{q}\cdot p) \, \insertleg{p}{q}{n}\, \cC(1,2,\ldots,n) =
(\epsilon_{q}\cdot p)\,\sum_{i=1}^{n-1}\, (\epsilon_{p}\cdot x_{i})\, 
 \insertleg{p}{q}{n}\, \cC(1,2,\ldots,i,p,i{+}1,\ldots,n)
\, .
\label{jan25}
\eeq
This expression is ready to be recast into the Parke-Taylor form by moving leg $p$ in the factors of $\cC(1,\ldots,i,p,i{+}1,\ldots,n)$ next to leg $n$ such that the numerator $\sigma_{p,n}$ gets cancelled. This can be achieved by means of Kleiss-Kuijf (KK) relations \cite{Kleiss:1988ne, DelDuca:1999rs}
\be
\cC(1,A, n, B) = (-)^{|B|}\, \sum_{\sigma\in A \shuffle B^{t}}
\cC(1,\sigma,n) \ .
\label{shufflerel}
\ee
The shuffle product of the sets $A \equiv \{ \alpha_1,\alpha_2,\ldots,\alpha_{|A|}\}$ and $B\equiv \{ \beta_1,\beta_2,\ldots,\beta_{|B|}\}$ in (\ref{shufflerel}) (with cardinality $|A|$, $|B|$ and reversal $B^{t}\equiv \{ \beta_{|B|},\ldots,\beta_2,\beta_1\}$) is defined recursively via
\be
\emptyset\shuffle A=  A \shuffle\emptyset = A ,\qquad
A\shuffle B \equiv \{\alpha_1(\alpha_2 \ldots \alpha_{|A|} \shuffle B)\} + \{\beta_1(\beta_2 \ldots \beta_{|B|}
\shuffle A )\}
\label{defshuffle}
\ee
and amounts to summing all permutations of $A \cup B$ which preserve the individual orderings of $A$ and $B$.
Applying this identity to \eqn{jan25} we may rewrite the last term as
\be
\cC(1,2,\ldots,i,p,i{+}1, \ldots, n) =
(-)^{n-i-1} \sum_{\sigma\in \{1,2,\ldots,i\} \atop{\shuffle \{n-1,n-2,\ldots,i+1\}} }
\cC(\sigma ,p,n)\, ,
\label{KKsign}
\ee
which leads to the following combination of Parke-Taylor factors,
\beq
C_{pp}'\, (\epsilon_{q}\cdot p) \, \insertleg{p}{q}{n}\, \cC(1,2,\ldots,n) =
(\epsilon_{q}\cdot p)\, \sum_{i=1}^{n-1} (-)^{n-i-1} \, (\epsilon_{p}\cdot x_{i})\, 
\sum_{\sigma\in \{1,2,\ldots,i\} \atop{\shuffle \{n-1,n-2,\ldots,i+1\}} }
\cC(\sigma ,p,q,n)
\, .
\label{jan25a}
\eeq
There is still freedom to simplify (\ref{jan25a}) using additional KK relations (\ref{shufflerel}). The particularly economic representation
\beq
\insertleg{p}{q}{n}\, \cC(1,2,\ldots,i,p,i{+}1,\ldots,n) = - \sum_{\sigma \in \{q\} \atop \shuffle \{1,2,\ldots,i\}} \cC(\sigma,p,i{+}1,\ldots,n) \ ,
\label{jan25b}
\eeq
can be conveniently verified in an ${\rm SL}(2,\mathbb C)$ frame where $\sigma_n \rightarrow \infty$. This yields the following compact alternative to (\ref{jan25a}):
\be
C_{pp}'\, (\epsilon_{q}\cdot p) \, \insertleg{p}{q}{n}\, \cC(1,2,\ldots,n)= - (\epsilon_{q}\cdot p)
\sum_{i=1}^{n-1} \, (\epsilon_{p}\cdot x_{i})
 \sum_{\sigma\in \{q\} \atop{\shuffle \{1, 2,\ldots, i\}}} 
\cC(\sigma, p, i{+}1, \ldots, n)\, .
\ee
The second cross-term in (\ref{2gmasteralt}) with $p$ and $q$ swapped can be addressed in the same manner. 

\subsection{Two-graviton contributions $(\epsilon_{p}\cdot  \epsilon_{q})$}
\label{sec43}

Moving on to the last term in \eqn{2gmasteralt}, one can use the peculiar cross-ratio identity of \cite{Cardona:2016gon}, 
\be
\frac{s_{pq}}{\sigma^{2}_{p,q}} = \sum_{i\neq a,p,q}s_{pi}\, \frac{\sigma_{i,a}}{\sigma_{i,p}
\, \sigma_{p,q}\, \sigma_{q,a}}\, , \ \ \ \ \ a \in \{1,2,\ldots,n\} \ ,
\label{crossrel}
\ee
which holds in the presence of momentum conservation and the scattering equations. Once the accompanying Parke-Taylor factors ${\cal C}(1,2,\ldots,n)$ are expanded in a KK-basis of ${\cal C}(\ldots,i,a)$ via (\ref{shufflerel}), any term on the right hand side of (\ref{crossrel}) can be brought into the desired form (\ref{desired}).

 Alternatively, one can simply apply the scattering equations in a frame where $\sigma_n \rightarrow \infty$ and replace $\frac{s_{pq}}{\sigma_{p,q}}=\sum_{i=1}^{n-1} \frac{ s_{ip}}{\sigma_{i,p}}$. This choice allows to rewrite the last term in \eqn{2gmasteralt} as
\beq
(\epsilon_{p}\cdot\epsilon_{q})\,
\frac{s_{pq}}{\sigma^{2}_{p,q}} \,\cC(1,2,\ldots,n) = (\epsilon_{p}\cdot\epsilon_{q})\,
\sum_{i=1}^{n-1} \, s_{ip}\,
 \sum_{\sigma\in\{q,p\} \atop{ \shuffle \{1,2, \ldots, i{-}1 \}}}
 \cC(\sigma,i, \ldots, n) \, .
\eeq
We could have applied the cross relation \eqn{crossrel} with $p$
and $q$ swapped, and the equality of the resulting color-ordered gluon amplitudes
follows from the BCJ relations \cite{Bern:2008qj},
\beq
\sum_{i=1}^{n-1} \, s_{ip}\,
 \sum_{\sigma\in\{q,p\} \atop{ \shuffle \{1, 2,\ldots, i{-}1 \}}}
 {\cal A}(\sigma,i, \ldots, n)
 = \sum_{i=1}^{n-1} \, s_{iq}\,
 \sum_{\sigma\in\{p,q\} \atop{ \shuffle \{1,2, \ldots, i{-}1 \}}}
 {\cal A}(\sigma,i, \ldots, n) \ .
\eeq
Hence, we may as well symmetrize the above result in $p$ and $q$.

\subsection{Amplitude relations for $n$ gluons and two gravitons}

By assembling the results from the previous subsections, (\ref{2gmasteralt}) yields the following final result for the two-graviton case
\begin{align} 
&\cA_{\text{EYM}} (1, \ldots,n; p,q) 
=
\sum_{1=i\leq j}^{n-1}   (\epsilon_{p}\cdot x_{i})\, (\epsilon_{q}\cdot x_{j})\, 
\cA(1,\ldots,i,p,i+1,\ldots, j, q,j+1,\ldots, n)   \label{twograv} \\
&\  - \sum_{i=1}^{n-1}  \Big\{\,
(\epsilon_{p} \! \cdot \! x_{i})\, (\epsilon_{q} \! \cdot \! p) 
 \! \! \! \sum_{\sigma\in \{q\} \atop{\shuffle \{1, \ldots, i\}}} \! \! \! 
\cA(\sigma, p, i{+}1, \ldots, n)
+ 
{ s_{pi} \, (\epsilon_{p} \! \cdot \! \epsilon_{q}) \over 2}   \! \! \! \! \!
 \sum_{\sigma\in\{ q, p\} \atop{ \shuffle \{1, \ldots, i{-}1\}}} \! \! \! \! \!
\cA( \sigma, i, i{+}1, \ldots ,n) 
 \, \Big\} + (  p \leftrightarrow q) \, ,
\notag
\end{align}
where the symmetrization over $p$ and $q$ applies to both lines of (\ref{twograv}). This result expresses a two-graviton 
and $n$-gluon amplitude through $(n{+}2)$-point pure gluon amplitudes and agrees with (\ref{twogravintro}) after expanding the sum over shuffles. The result derived from the CHY formula is valid 
in any dimension. The simplest non-trivial examples involve two and three gluons, respectively,
\begin{align}
{\cal A}_{\rm EYM}(1, 2; 3, 4) &=
(\ep_3\cdot k_4)(\ep_4 \cdot x_1) {\cal A}(1,2,3,4)
+(\ep_4\cdot k_3)(\ep_3 \cdot x_1) {\cal A}(1,2,4,3) \notag \\
&
-(\ep_3\cdot x_1)(\ep_4 \cdot x_1) {\cal A}(1,3,2,4)
- s_{13} (\ep_3 \cdot \ep_4) {\cal A}(1,2,4,3) \, , \label{ex2plus2}\\
{\cal A}_{\rm EYM}(1, 2,3; 4, 5) &=  (\ep_4 \cdot x_2) (\ep_5\cdot k_4)\,
{\cal A}( 1, 2, 4, 5, 3) 
  + (\ep_4 \cdot x_2) (\ep_5 \cdot x_1) \,  {\cal A}(1, 5, 2, 4, 3) \notag \\
  &+   (\ep_4 \cdot x_1) (\ep_5 \cdot x_1)\, {\cal A}(1, 4, 5, 2, 3)
  + (\ep_4 \cdot x_2) (\ep_5 \cdot x_2) \, {\cal A}(1, 2, 4, 5, 3)  \notag \\
  &
  - (\ep_4 \cdot x_1) (\ep_5 \cdot k_4)\, \big[ {\cal A}(1, 5, 4, 2, 3)+ {\cal A}(5, 1, 4, 2, 3) \big]  \notag \\
 &+ \tfrac{1}{2} (\ep_4\cdot \ep_5) \big[ s_{24}  {\cal A}( 1, 3, 5, 4, 2) - s_{14}  {\cal A} ( 1, 2, 3,5, 4) \big] + (4\leftrightarrow 5) \ ,
 \label{ex3plus2}
 \end{align}
 where gauge invariance under $\ep_p \rightarrow p$ can be checked via BCJ relations \cite{Bern:2008qj} among the ${\cal A}(\ldots)$. In terms of the only BCJ-independent four-point amplitude ${\cal A}(1,2,3,4)$, (\ref{ex2plus2}) can be brought into the manifestly gauge invariant form
\begin{align}
{\cal A}_{\rm EYM}(1, 2; 3, 4) &=
{\cal A}(1,2,3,4) \times \Big\{ (\ep_3\cdot k_4)(\ep_4 \cdot k_1)  + \frac{ s_{23}}{s_{13}}(\ep_4\cdot k_3)(\ep_3 \cdot k_1) 
 \label{gaugeinv} \\
& \ \ \ \ \ \ \ \ \ \ \ \ \ \ \ \ \ \ \ \ \ \ \ \ \ \ \ \   - \frac{ s_{12}}{s_{13}}(\ep_3\cdot k_1)(\ep_4 \cdot k_1) 
- s_{23} (\ep_3 \cdot \ep_4) \Big\}
\notag
  \ ,
\end{align}
and similar expressions can be found for ${\cal A}_{\rm EYM}(1, 2,3; 4, 5)$ by reducing the right hand side 
of (\ref{ex3plus2}) to a five-point BCJ basis such as $\{{\cal A}(1,2,3,4,5),{\cal A}(1,3,2,4,5)\}$.

\section{Three gravitons}
\label{sec5}

We proceed to the amplitudes with three gravitons, where the relevant Pfaffian $\Psi_{r=3}$ is given as
\begin{align}
\text{Pf} \, \Psi_{r=3} &= C_{pp}C_{qq}C_{rr} + \Big[ C_{pp} \frac{ (\ep_q \cdot r)(\ep_r\cdot q) - s_{qr} (\ep_q \cdot \ep_r)}{\sigma_{q,r}^2} + {\rm cyc}(p,q,r) \Big] \notag \\
&+\frac{1}{\sigma_{p,q} \sigma_{q,r} \sigma_{r,p} } \Big[ (\ep_p \cdot q)(\ep_q \cdot r)(\ep_r \cdot p)-(\ep_p \cdot r)(\ep_r \cdot q)(\ep_q \cdot p) \Big] \label{threeg1}
\\
&+\frac{1}{\sigma_{p,q} \sigma_{q,r} \sigma_{r,p} } \Big[ (\ep_p \cdot \ep_q) \{ s_{pr} (\ep_r\cdot q) - s_{qr} (\ep_r\cdot p) \} + {\rm cyc}(p,q,r) \Big] \, , \notag
\end{align}
where ${\rm cyc}(p,q,r)$ instructs to add the two cyclic permutations $(p,q,r) \rightarrow (q,r,p)$
as well as $(p,q,r) \rightarrow (r,p,q)$.
As in the case of two gravitons, it is convenient to write $C_{pp}$ as
\ba 
C_{pp} = C'_{pp} + (\epsilon_p \cdot q) \, { \sigma_{q,n} \over  \sigma_{q,p} \sigma_{p,n} } +
(\epsilon_p \cdot r)  \, { \sigma_{r,n} \over  \sigma_{r,p} \sigma_{p,n} }  \, ,
\ea
with $C'_{pp}$ as given in (\ref{eq:cprime}) and an analogous splitting of $C_{qq}$ and $C_{rr}$. Similar to the absence of tensor structures $(\ep_p\cdot q )(\ep_q\cdot p)$ in the two-graviton case, we note the vanishing of  three classes of terms in (\ref{threeg1}) given by
\beq
(\ep_p \cdot q)(\ep_q \cdot r)(\ep_r \cdot p) \ , \ \ \ \ \ \ (\ep_p \cdot q)(\ep_q \cdot p)(\ep_r \cdot p)
\ , \ \ \ \ \ \ (\ep_p \cdot q)(\ep_q \cdot p)(\ep_r \cdot x_j) 
\label{threeg2}
\eeq
and their permutations in $p,q,r$. The non-vanishing contributions to (\ref{threeg1}) may be organized according to the contractions of polarization vectors and momenta. It turns out that there are five such independent tensor structures, and we will discuss them one by one. 


\subsection{Three-graviton contributions $(\ep_p \cdot \ep_q)(\ep_r \cdot q)$}
\label{sec51}

We begin with the term proportional to $(\ep_p \cdot \ep_q)(\ep_r \cdot q)$, where the corresponding $\sigma$-dependence stems from the first and the third line of (\ref{threeg1}),
\beq
\text{Pf} \, \Psi_{r=3} \, \Big|_{ (\ep_p \cdot \ep_q)(\ep_r \cdot q) } = \frac{ s_{pq} \sigma_{q,n} }{\sigma_{r,q} \sigma_{r,n} \sigma_{p,q}^2} - \frac{ s_{pr}}{\sigma_{p,q} \sigma_{q,r} \sigma_{p,r}}
\, .
\eeq
In a frame where $\sigma_n \rightarrow \infty$, these two terms can be combined through a single scattering equation $\frac{s_{rp}}{\sigma_{r,p}}+\frac{s_{qp}}{\sigma_{q,p}}= \sum_{j=1}^{n-1} \frac{ s_{pj}}{\sigma_{p,j}}$. In combination with an $n$-particle Parke-Taylor factor, KK-rearrangements similar to those in section \ref{sec4} yield the ${\rm SL}(2,\mathbb C)$-covariant result
\begin{align}
{\cal C}(1,2,\ldots,n) \,\text{Pf} \, \Psi_{r=3} \, \Big|_{ (\ep_p \cdot \ep_q)(\ep_r \cdot q) } =  \sum_{j=1}^{n-1} s_{jp}\sum_{\sigma \in \{r,q,p\} \atop{\shuffle \{ 1,2,\ldots, j-1\}} } {\cal C}(\sigma,j,j+1,\ldots,n)
\end{align}
in terms of $(n{+}3)$-particle Parke-Taylor factors.

\subsection{Three-graviton contributions $(\ep_p \cdot \ep_q)  (\ep_r \cdot x_j) $}

We move on to tensor structures $(\ep_p \cdot \ep_q)  (\ep_r \cdot x_j)$ stemming from the end of the first line of (\ref{threeg1}), 
\beq
\text{Pf} \, \Psi_{r=3} \, \Big|_{(\ep_p \cdot \ep_q)  (\ep_r \cdot x_j)  } = - \frac{ s_{pq} }{\sigma_{p,q}^2} \frac{ \sigma_{j,j+1} }{\sigma_{j,r} \sigma_{r,j+1}} \ .
\eeq
The techniques of section \ref{sec43} for the Parke-Taylor factor ${\cal C}(1,2,\ldots,j,r,j{+}1,\ldots,n)$
yield
\begin{align}
&{\cal C}(1,2,\ldots,n) \,\text{Pf} \, \Psi_{r=3} \, \Big|_{ (\ep_p \cdot \ep_q)  (\ep_r \cdot x_j) }= - \frac{ s_{pq} }{\sigma_{p,q}^2} \,{\cal C}(1,2,\ldots,j,r,j{+}1,\ldots,n) \notag \\
& =  -   \Big\{   \sum_{i=1}^j s_{ip} \! \! \! \! \!  \! \! \sum_{\sigma \in \{q,p\} \atop{\shuffle \{ 1,2,\ldots, i{-}1\}} } \! \! \! \! \! \! \! {\cal C}(\sigma,i,i{+}1,\ldots,j,r,j{+}1,\ldots,n)
  \label{3gb} \\
& \ \ \ \ \ \ +s_{pr} \sum_{\sigma \in \{q,p\} \atop{\shuffle \{ 1,2,\ldots, j\}} } {\cal C}(\sigma,r,j{+}1,\ldots,n)+ \sum_{i=j+1}^{n-1} s_{ip} \! \! \! \! \!\! \! \! \! \! \sum_{\sigma \in \{q,p\} \atop{\shuffle \{ 1,\ldots,j,r,j{+}1,\ldots i-1\}} }\! \! \! \! \! \! \! \! \! \! {\cal C}(\sigma,i,i{+}1,\ldots,n) \Big\} \ ,
\notag
\end{align}
where the hidden symmetry under $p\leftrightarrow q$ can be checked via BCJ relations.

\subsection{Three-graviton contributions $(\ep_p \cdot (q+r)) ( \ep_q \cdot r) (\ep_r \cdot x_j)$}

We jointly discuss the tensor structures $(\ep_p \cdot q) ( \ep_q \cdot r) (\ep_r \cdot x_j)$ and $(\ep_p \cdot r) ( \ep_q \cdot r) (\ep_r \cdot x_j)$ which arise from $C_{pp}\, C_{qq} \, C'_{rr}$,
\begin{align}
\text{Pf} \, \Psi_{r=3} \, \Big|_{ (\ep_p \cdot q) ( \ep_q \cdot r) (\ep_r \cdot x_j)  } &= \frac{ \sigma_{r,n} }{\sigma_{p,q} \sigma_{p,n} \sigma_{q,r} } \frac{ \sigma_{j,j+1} }{\sigma_{j,r} \sigma_{r,j+1} }
\\
\text{Pf} \, \Psi_{r=3} \, \Big|_{ (\ep_p \cdot r) ( \ep_q \cdot r) (\ep_r \cdot x_j)  } &= \frac{ \sigma_{r,n}^2 }{\sigma_{p,r} \sigma_{p,n} \sigma_{q,r}\sigma_{q,n} } \frac{ \sigma_{j,j+1} }{\sigma_{j,r} \sigma_{r,j+1} } \ .
\end{align}
The Schouten identity $\sigma_{r,n} \sigma_{p,q} = \sigma_{r,p}\sigma_{n,q} - \sigma_{r,q}\sigma_{n,p}$ and the frame-choice $\sigma_{n}\rightarrow \infty$ are helpful to combine these expressions with a Parke-Taylor factor,
\begin{align}
{\cal C}(1,2,\ldots,n)\text{Pf} \, \Psi_{r=3} \, \Big|_{ (\ep_p \cdot q) ( \ep_q \cdot r) (\ep_r \cdot x_j)  } &=  \sum_{\sigma \in \{p,q\} \atop{\shuffle \{ 1,2,\ldots, j\}} } {\cal C}(\sigma,r,j{+}1,\ldots,n)
\label{3dA}
\\
{\cal C}(1,2,\ldots,n)\text{Pf} \, \Psi_{r=3} \, \Big|_{ (\ep_p \cdot r) ( \ep_q \cdot r) (\ep_r \cdot x_j)  } &= \sum_{\sigma \in \{p,q\} \atop{\shuffle \{ 1,2,\ldots, j\}} } {\cal C}(\sigma,r,j{+}1,\ldots,n) + (p\leftrightarrow q)
 \ .
 \label{3dB}
\end{align}
Distributing the two terms in (\ref{3dB}) among two permutations of (\ref{3dA}) yields six terms of the form
\begin{align}
&{\cal C}(1,2,\ldots,n)\text{Pf} \, \Psi_{r=3} \, \Big|_{ (\ep_p \cdot (q+r)) ( \ep_q \cdot r) (\ep_r \cdot x_j)  }   =  \sum_{\sigma \in \{p,q\} \atop{\shuffle \{ 1,2,\ldots, j\}} } {\cal C}(\sigma,r,j{+}1,\ldots,n)   \ .
    \end{align}

\subsection{Three-graviton contributions $( \ep_r \cdot p)  (\ep_p \cdot x_i)  (\ep_q \cdot x_j)$}

The tensor structure $( \ep_r \cdot p)  (\ep_p \cdot x_i)  (\ep_q \cdot x_j)$ due to $C_{rr} \, C'_{pp} \,C'_{qq}$ is accompanied by
\beq
\text{Pf} \, \Psi_{r=3} \, \Big|_{ ( \ep_r \cdot p)  (\ep_p \cdot x_i)  (\ep_q \cdot x_j) } = \frac{ \sigma_{ p,n} }{\sigma_{p,r} \sigma_{r,n}} \frac{ \sigma_{i,i+1} \sigma_{j,j+1}}{\sigma_{i,p} \sigma_{p,i+1} \sigma_{j,q} \sigma_{q,j+1}} \ .
\eeq
There are three cases to consider for the relative positions of $i$ and $j$,
\beq
{\cal C}(1,\ldots,n)\text{Pf} \, \Psi_{r=3} \, \Big|_{ ( \ep_r \cdot p)  (\ep_p \cdot x_i)  (\ep_q \cdot x_j) } = 
\left\{ \begin{array}{cl}\displaystyle
- \! \! \! \!  \sum_{\sigma \in \{r\} \atop{\shuffle \{ 1,2,\ldots, i\}} }  \! \! \! \! {\cal C}(\sigma,p,i{+}1,\ldots,j,q,j{+}1,\ldots,n)  &: \ i<j \\
\displaystyle
- \! \! \! \! \!  \sum_{\sigma \in \{r\} \atop{\shuffle \{ 1,\ldots, j,q,j{+}1,\ldots,i\}} }  \! \! \! \! \!  {\cal C}(\sigma,p,i{+}1,\ldots,n)  &:\  i>j \\
%
\end{array} \right.
\label{3e2} 
\eeq
as well as the case where $i=j$,
\beq
{\cal C}(1,\ldots,n)\text{Pf} \, \Psi_{r=3} \, \Big|_{ ( \ep_r \cdot p)  (\ep_p \cdot x_i)  (\ep_q \cdot x_i) } = - \! \! \! \!  \! \sum_{\sigma \in \{r\} \atop{\shuffle \{ 1,2,\ldots, i\}} }  \! \! \! \! \! {\cal C}(\sigma,p,q,j{+}1,\ldots,n)  - \! \! \! \! \! \!  \sum_{\sigma \in \{r\} \atop{\shuffle \{ 1,2,\ldots, i,q\}} }  \! \! \! \! \! \!  {\cal C}(\sigma,p,j{+}1,\ldots,n)  \ .
\label{3e1} 
\eeq
Note that summing over all choices of $i,j$ combines (\ref{3e2}) and (\ref{3e1}) to
\begin{align}
&C'_{pp} \, C'_{qq}\,  {\cal C}(1,\ldots,n) \, \frac{  \sigma_{p,n} }{\sigma_{p,r} \sigma_{r,n}}
= -  \Big\{
\sum_{1=j\leq i}^{n-1}  (\ep_p \cdot x_i)  (\ep_q \cdot x_j) \! \! \! \! \! \sum_{\sigma \in \{r\} \atop{\shuffle \{ 1,\ldots,j,q,j+1,\ldots, i\}} } \! \! \! \! \! {\cal C}(\sigma,p,i{+}1,\ldots,n) 
  \notag \\
& \ \ \ \ \ \ \ \ \ \ \ \ \ \ \   \ \ \ \ \ + \sum_{1=i\leq j}^{n-1}  (\ep_p \cdot x_i)  (\ep_q \cdot x_j)  \sum_{\sigma \in \{r\} \atop{\shuffle \{ 1,2,\ldots, i\}} } {\cal C}(\sigma,p,i{+}1,\ldots,j ,q,j{+}1,\ldots,n) 
\Big\} \ .
    \end{align}  

\subsection{Three-graviton contributions $ (\ep_p \cdot x_i)  (\ep_q \cdot x_j)(\ep_r \cdot x_k)$}
\label{sec55}

Finally, the term $C'_{pp} \, C'_{qq} \, C'_{rr} $ contributes tensor structures $(\ep_p \cdot x_i)  (\ep_q \cdot x_j) ( \ep_r \cdot x_k) $ along with
\beq
\text{Pf} \, \Psi_{r=3} \, \Big|_{  (\ep_p \cdot x_i)  (\ep_q \cdot x_j) ( \ep_r \cdot x_k)  } =  \frac{ \sigma_{i,i+1} \sigma_{j,j+1} \sigma_{k,k+1}}{\sigma_{i,p} \sigma_{p,i+1} \sigma_{j,q} \sigma_{q,j+1} \sigma_{k,r} \sigma_{r,k+1}}  \ ,
\eeq
where clashes among the summation variables $i,j,k$ require applications of the Schouten identity:
\begin{align}
&{\cal C}(1,\ldots,n) \,
\text{Pf} \, \Psi_{r=3} \, \Big|_{  (\ep_p \cdot x_i)  (\ep_q \cdot x_j) ( \ep_r \cdot x_k)  } \\
& \ = \left\{\begin{array}{cl}
 {\cal C}(1,2,\ldots,i,p,i{+}1,\ldots,j,q,j{+}1,\ldots,k,r,k{+}1,\ldots,n) &: \ i<j<k \\
  {\cal C}(1,2,\ldots,i,p,q,i{+}1,\ldots,k,r,k{+}1,\ldots,n)  + (p\leftrightarrow q)&: \ i=j<k \\
   {\cal C}(1,2,\ldots,i,p,q,r,i{+}1,\ldots,n) +{\rm perm}(p,q,r) &: \ i=j=k \\
\end{array} \right. \notag 
\end{align}
The above cases and their permutations in $p,q,r$ can be combined to
\begin{align}
& C'_{pp} \, C'_{qq} \, C'_{rr}  \, {\cal C}(1,\ldots,n)  = 
\sum_{1=i\leq j\leq k}^{n-1}  (\ep_p \cdot x_i)  (\ep_q \cdot x_j)(\ep_r \cdot x_k) \\
& \ \ \ \ \times  {\cal C}(1,2,\ldots,i,p,i{+}1,\ldots,j,q,j{+}1,\ldots,k,r,k{+}1,\ldots,n)  +{\rm perm}(p,q,r) \ .  \notag
\end{align} 

\subsection{Amplitude relations for $n$ gluons and three gravitons}

Assembling all the terms from the above sections \ref{sec51} to \ref{sec55} yields the following amplitude relation
\begin{align}
 &\! \! \! \cA_{\text{EYM}} (1,2, \ldots,n; p,q,r) = (\ep_p \cdot \ep_q)(\ep_r \cdot q) \sum_{j=1}^{n-1} s_{jp}\sum_{\sigma \in \{r,q,p\} \atop{\shuffle \{ 1,2,\ldots, j-1\}} } {\cal A}(\sigma,j,j{+}1,\ldots,n) \notag \\
&-\, \frac{ (\ep_p \cdot \ep_q)}{2} \sum_{j=1}^{n-1}  (\ep_r \cdot x_j)  \Big\{
\sum_{i=1}^j s_{ip} \! \! \! \! \! \sum_{\sigma \in \{q,p\} \atop{\shuffle \{ 1,2,\ldots, i-1\}} } \! \! \! \! \! {\cal A}(\sigma,i,i{+}1,\ldots,j,r,j{+}1,\ldots,n)
\notag  \\
& \ \ \ \ \ \  \ \ \ \ \ \  + s_{pr} \sum_{\sigma \in \{q,p\} \atop{\shuffle \{ 1,2,\ldots, j\}} } {\cal A}(\sigma,r,j{+}1,\ldots,n)
+ \sum_{i=j+1}^{n-1} s_{ip} \! \! \! \! \!\! \! \! \! \! \sum_{\sigma \in \{q,p\} \atop{\shuffle \{ 1,\ldots,j,r,j{+}1,\ldots i-1\}} }\! \! \! \! \! \! \! \! \! \! {\cal A}(\sigma,i,i{+}1,\ldots,n) \Big\} \notag \\
& -  ( \ep_r \cdot p) \Big\{
\sum_{1=j\leq i}^{n-1}  (\ep_p \cdot x_i)  (\ep_q \cdot x_j) \! \! \! \! \! \sum_{\sigma \in \{r\} \atop{\shuffle \{ 1,\ldots,j,q,j{+}1,\ldots, i\}} } \! \! \! \! \! {\cal A}(\sigma,p,i{+}1,\ldots,n) 
  \label{threegr} \\
& \ \ \ \ \ \  \ \ \ \ \ \ + \sum_{1=i\leq j}^{n-1}  (\ep_p \cdot x_i)  (\ep_q \cdot x_j)  \sum_{\sigma \in \{r\} \atop{\shuffle \{ 1,2,\ldots, i\}} } {\cal A}(\sigma,p,i{+}1,\ldots,j ,q,j{+}1,\ldots,n) 
\Big\} \notag \\
&+\sum_{1=i\leq j\leq k}^{n-1}  (\ep_p \cdot x_i)  (\ep_q \cdot x_j)(\ep_r \cdot x_k)    {\cal A}(1,2,\ldots,i,p,i{+}1,\ldots,j,q,j{+}1,\ldots,k,r,k{+}1,\ldots,n)\notag \\
& + (\ep_p \cdot (q{+}r)) ( \ep_q \cdot r) \sum_{j=1}^{n-1}  (\ep_r \cdot x_j)  \sum_{\sigma \in \{p,q\} \atop{\shuffle \{ 1,2,\ldots, j\}} } {\cal A}(\sigma,r,j{+}1,\ldots,n)  +{\rm perm}(p,q,r)  \ ,\notag
\end{align}
where the symmetrization over the three gravitons with momenta $p,q,r$ applies to all the lines. 
The simplest non-trivial example of (\ref{threegr}) involves two gluons and three gravitons
\begin{align}
A_{\rm EYM}(1, 2; 3, 4,5) &= (\ep_3 \cdot x_1)(\ep_4 \cdot x_1) (\ep_5 \cdot x_1)
 {\cal A}(1, 3, 4, 5, 2)  
+ (\ep_5\cdot k_3) (\ep_3 \cdot x_1)(\ep_4 \cdot x_1) {\cal A}( 1, 4, 2,5, 3)  \notag \\
&\! \! \! \! \! -  (\ep_4 \cdot (k_3+k_5)) (\ep_5 \cdot k_3)(\ep_3 \cdot x_1) {\cal A}( 1, 2, 4, 5, 3) 
+ (\ep_3 \cdot \ep_5) (\ep_4 \cdot k_5)  s_{13}  {\cal A}( 1, 2, 4, 5, 3)  \notag \\
 &\! \! \! \! \!  +\tfrac{1}{2} (\ep_4\cdot \ep_5)(\ep_3 \cdot x_1)  
 \big[ s_{3 4} {\cal A}( 3, 1, 2,5, 4)  - s_{1 4} {\cal A} (1, 3, 2, 5, 4)  \big] + {\rm perm}(3,4,5) \ ,
  \end{align}
where gauge invariance under $\ep_p \rightarrow p$ can be easily checked by casting all partial amplitudes on the right hand side into a two-element BCJ-basis.

\section{Four and more gravitons}
\label{sec6}

It is straightforward to extend the results in the previous sections to the case with four gravitons and a single-trace contribution of gluons. As in the case of two and three  gravitons, we again write $C_{pp}$ as
\ba 
C_{pp} = C'_{pp} + (\epsilon_p \cdot q) \, { \sigma_{q,n} \over  \sigma_{q,p} \sigma_{p,n} } +
(\epsilon_p \cdot r)  \, { \sigma_{r,n} \over  \sigma_{r,p} \sigma_{p,n} }  +
(\epsilon_p \cdot t)  \, { \sigma_{t,n} \over  \sigma_{t,p} \sigma_{p,n} }  \, ,
\ea
with $C'_{pp}$ as given in (\ref{eq:cprime}) and an analogous splitting of $C_{qq}$, $C_{rr}$ and $C_{tt}$. Spelling out the complete four-graviton Pfaffian of $\Psi_{r=4}$ \cite{Cachazo:2014nsa} is a tedious but straightforward generalization of (\ref{threeg1}). Using these definitions we can proceed just as in the three-graviton case and identify the following $16$ permutation independent tensor structures that do not cancel:
\[
\begin{array}{llll}
C'_{pp} C'_{qq} (\ep_r \cdot q) (\ep_t \cdot p) 
& C'_{pp} (\ep_q \cdot t) (\ep_r \cdot p)(\ep_t \cdot p)
& C_{tt}' (\ep_q \cdot \ep_r) (\ep_p \cdot q) 
& (\ep_p \cdot \ep_q)(\ep_r\cdot \ep_t) 
\\
(\ep_p \cdot \ep_t) (\ep_q\cdot t)(\ep_r \cdot p)
& C'_{pp} (\ep_q \cdot p) (\ep_r \cdot p) (\ep_t \cdot p)
& C'_{pp} (\ep_r \cdot \ep_t) (\ep_q \cdot p)
& C'_{pp}C'_{qq} (\ep_r \cdot \ep_t) 
\\
(\ep_q \cdot \ep_t) (\ep_p \cdot t)(\ep_r \cdot p)
&C_{pp}' (\ep_q \cdot r) (\ep_r \cdot t)(\ep_t \cdot p)
&C_{pp}' C_{tt}' (\ep_q \cdot r) (\ep_r \cdot p)
&C'_{pp} C_{rr}' C_{tt}' (\ep_q \cdot p)  
\\
(\ep_p \cdot \ep_t) (\ep_q \cdot p) (\ep_r \cdot p) \ 
&C'_{qq} (\ep_p \cdot q) (\ep_r \cdot p)(\ep_t \cdot p) \ 
&C_{pp}' C_{tt}' (\ep_q \cdot p) (\ep_r \cdot p) \
&C'_{pp} C_{qq}' C_{rr}' C_{tt}' .
\label{eq:fourgravitonlist}
\end{array}
\]
The previous techniques and results for dealing with two and three gravitons can be recycled to relate all the above tensor structures to YM subamplitudes, the expressions being of course too long to be displayed in this work. The most nontrivial case $\sim (\ep_p \cdot \ep_q)(\ep_r\cdot \ep_t) $ is discussed in appendix \ref{app4G}.

Also in cases with any number of gravitons, $\textrm{SL}(2,\mathbb C)$-invariance requires the CHY integrands to be built from products of Parke-Taylor factors and cross-ratios which possibly lead to higher-order poles in $\sigma_{i,j}$.  According to \cite{Cardona:2016gon}, any such CHY integrand can always be reduced to linear combinations of single-cycle Parke-Taylor factors which signal the single-trace color ordered amplitudes in YM. Hence, it follows that the above procedure can in principle be generalized to any number of gravitons.

\section{Towards multitrace contributions} \label{section:multitrace}

In the following two sections, we provide an indication of how the techniques of this work also apply to multitrace contributions to EYM amplitudes. Note that these results apply to EYM extended by B-field and dilaton couplings \cite{Cachazo:2014nsa, Cachazo:2014xea} reflecting their string theoretic underpinning. While an exhaustive discussion is relegated to future work, we will consider the two particularly simple examples of double-trace amplitudes involving gluons only as well as those with one single graviton.

In order to lighten the notation, we strip off the ubiquitous $n$-particle Pfaffian from the subsequent $n$-gluon integrands
\beq
{\cal I}^{\rm EYM}_{\{r,n-r\}}(1,2,\ldots,r \, | \, r{+}1,\ldots,n) \equiv {\cal J}^{\rm EYM}_{\{r,n-r\}}(1,2,\ldots,r \, | \, r{+}1,\ldots,n) \cdot \text{Pf}'\,  \Psi_{n}( \{k_a,\epsilon,\sigma\})
\label{light}
\eeq
and focus on the reduced integrands ${\cal J}_{\{r,n-r\}}^{\rm EYM}$ on the right hand side. The subscript $\{r,n-r\}$ refers to having $r$ and $n{-}r$ gluons in the first and second trace, respectively. Similarly, the reduced double-trace integrands ${\cal J}_{\{r,n-r\}+1}^{\rm EYM}$ for $n$ gluons and one graviton to be discussed in section \ref{section:moremultitrace} are understood to exclude the overall $(n{+}1)$-particle Pfaffian.

\subsection{Double-trace amplitude relations without gravitons}

The CHY integrand for double-trace contributions to gluon amplitudes is given by\footnote{We are following the normalization conventions of \cite{Cachazo:2014xea}.} \cite{Cachazo:2014nsa, Cachazo:2014xea} 
\beq
{\cal J}^{\rm EYM}_{\{r,n-r\}}(1,2,\ldots,r \, | \, r{+}1,\ldots,n) = s_{12\ldots r} \, {\cal C}(1,2,\ldots,r)  \, {\cal C}(r{+}1,\ldots,n)  
\label{mtrace1}
\eeq
with multiparticle Mandelstam variables
\beq
s_{12\ldots r} \equiv \sum_{1 \leq i<j}^r (k_i \cdot k_j) \ .
\label{multimand}
\eeq
Using cross-ratio identities \cite{Cardona:2016gon} similar to (\ref{crossrel}), one can rewrite the product of Parke-Taylor factors in (\ref{mtrace1}) in terms of a single $n$-particle Parke-Taylor factor\footnote{The problem of evaluating CHY integrals involving multiple Parke-Taylor factors has been actively studied in the recent literature \cite{Cachazo:2015nwa, Gomez:2016bmv, Cardona:2016gon}. The string-theory analogue of this problem where scattering equations translate into integration by parts is relevant to reduce tree-level amplitudes of the open superstring \cite{Mafra:2011nv, Mafra:2011nw} and the open bosonic string \cite{Huang:2016tag} to an $(n-3)!$ basis of worldsheet integrals.}. This generalizes the procedure of section \ref{sec43} and reduces any double-trace subamplitude to linear combinations of their single-trace counterparts. In the remainder of this section, we will derive the following all-multiplicity formula:
\begin{align}
&{\cal A}_{\rm EYM}(1,2,\ldots,r \, | \, r{+}1,\ldots,n) = \sum_{i=1}^{r-1} \sum_{j=r+2}^n (-1)^{j-i} \, s_{ij}\! \! \! \! \! \!  \! \! \! \! \!\sum_{\rho \in \{1,2,\ldots,i-1\}  \atop{ \shuffle \{r-1,r-2,\ldots,i+1\}}}  \sum_{\tau \in \{{j+1} ,j+2,\ldots,n \} \atop{\shuffle \{ j-1, j-2,\ldots, r+2 \}}}  \! \! \! \! \! \! \! \! \! \! \! {\cal A}(\rho,i,j,\tau,r{+}1,r) 
\label{mtrace2}
\end{align}
For a small number of gluons in the second cycle $\{ r{+}1,\ldots,n\}$, say $n-r=2,3,4$,
the general expression (\ref{mtrace2}) simplifies to
\begin{align}
{\cal A}_{\rm EYM}(1,2,\ldots,r \, | \,p,q) &= \sum_{i=1}^{r-1} (-1)^{r-i} s_{iq}  \! \! \! \! \! \! \!   \!\sum_{\rho \in \{1,2,\ldots,i-1\}\atop{ \shuffle \{r-1,r-2,\ldots,i+1\}}} \! \!  \! \!  {\cal A}(\rho,i,q,p,r) 
\label{mtrace3} \\
{\cal A}_{\rm EYM}(1,2,\ldots,r \, | \,p,q,t) &= \sum_{i=1}^{r-1} (-1)^{r-i}   \! \! \! \! \! \!  \!  \sum_{\rho \in \{1,2,\ldots,i-1\} \atop{\shuffle \{r-1,r-2,\ldots,i+1\}}} \! \! \!  \! \big[ s_{iq}  {\cal A}(\rho,i,q,t,p,r) - s_{it}  {\cal A}(\rho,i,t,q,p,r)  \big]
\label{mtrace4} \\
{\cal A}_{\rm EYM}(1,2,\ldots,r \, | \,p,q,t,u) &= \sum_{i=1}^{r-1} (-1)^{r-i}   \! \! \! \! \! \! \!  \sum_{\rho \in \{1,2,\ldots,i-1\} \atop{\shuffle \{r-1,r-2,\ldots,i+1\}}} \! \! \!  \! \big[ s_{iq}  {\cal A}(\rho,i,q,t,u,p,r) + s_{iu}  {\cal A}(\rho,i,u,t,q,p,r) \notag \\
& \ \ \ \ \ \ \ \ \ \ \ \ \ \ \ \ \ \  -
 s_{it}  {\cal A}(\rho,i,t,q,u,p,r) -s_{it}  {\cal A}(\rho,i,t,u,q,p,r)   \big] \label{mtrace5}
\end{align}
with lowest-multiplicity examples
\begin{align}
{\cal A}_{\rm EYM}(1,2 \, | \, 3,4) &= -s_{14} {\cal A}(1, 2, 3, 4)
\\
{\cal A}_{\rm EYM}(1,2,3 \, | \, 4,5) &= s_{15} {\cal A}(2, 1, 5, 4, 3)  - s_{25} {\cal A}(1, 2, 5, 4, 3) 
\\
{\cal A}_{\rm EYM}(1,2,3,4 \, | \, 5,6) &=
   s_{26} \big[ {\cal A}(1, 3, 2, 6, 5, 4) + {\cal A}(3, 1, 2, 6, 5, 4) \big]  \notag \\
   & - s_{16} {\cal A}(3, 2, 1, 6, 5, 4) - s_{36} {\cal A}(1, 2, 3, 6, 5, 4) 
 \\
 {\cal A}_{\rm EYM}(1,2,3 \, | \, 4,5,6) &= 
s_{15} {\cal A}(2, 1, 5, 6, 4, 3)  - s_{16} {\cal A}(2, 1, 6, 5, 4, 3)  \notag \\
& - s_{25} {\cal A}(1, 2, 5, 6, 4, 3) + s_{26} {\cal A}(1, 2, 6, 5, 4, 3)  \ .
 \end{align}
Cyclicity within the individual traces and symmetry under exchange of the traces are non-manifest in these expressions but can be checked to hold via BCJ relations. Note that the integrands of open-string one-loop amplitudes
have been organized in terms of similar combinations of YM trees \cite{Mafra:2012kh} -- see in particular appendix B of \cite{Mafra:2014oia}. Hence, the above relations are expected to follow conveniently from the low-energy limit of one-loop diagrams of the type-I superstring.
 
\subsection{The derivation}
 
The derivation of (\ref{mtrace2}) is based on a more general form of the cross-ratio identity (\ref{crossrel}) \cite{Cardona:2016gon},
\beq
 -s_{r+1,r+2, \ldots ,n} = \sum_{i=1}^{r-1} \sum_{j=r+2}^{n} s_{ij} \, \frac{ \sigma_{j,r+1} \sigma_{i,r} }{\sigma_{i,j} \sigma_{r,r+1}} \ .
\label{mtrace6} 
\eeq
It holds in the presence of momentum conservation as well as scattering equations
and will be applied to the CHY integrand (\ref{mtrace1}) for different choices of the sets $\{r{+}1,r{+}2,\ldots,n\}$:
\begin{itemize}
\item For a cycle of length two, setting $(r{+}1,n) \rightarrow (p,q)$ yields
\begin{align}
{\cal J}^{\rm EYM}_{\{r,2\}}(1,2,\ldots,r \, | \, p,q)  &= - {\cal C}(1,2,\ldots,r) \times  \frac{1}{\sigma_{p,q}} \sum_{i=1}^{r-1} s_{iq} \frac{  \sigma_{i,r} }{\sigma_{i,q} \sigma_{r,p} }   \label{mtrace7}  \\
&= \sum_{i=1}^{r-1} s_{iq} (-1)^{r-i}   \! \! \! \! \! \!    \!\sum_{\rho \in \{1,2,\ldots,i-1\} \atop{ \shuffle \{r-1,\ldots,i+1\}}} \! \! \! \! \! \!   {\cal C}(\rho,i,q,p,r)  \ ,\notag 
\end{align}
which translates to the amplitude relation (\ref{mtrace3}). Here and in later cases, the numerator factor $\sigma_{i,r}$ in the first line
has been canceled after expanding the Parke-Taylor factor ${\cal C}(1,2,\ldots,r)$ of the $r$-particle cycle in a KK-basis of ${\cal C}(\ldots,i,r) \sim \sigma_{i,r}^{-1}$, see (\ref{shufflerel}).
\item For a cycle of length three, setting $(r{+}1,r{+}2,n) \rightarrow (p,q,t)$ yields
\begin{align}
{\cal J}^{\rm EYM}_{\{r,3\}}(1,2,\ldots,r \, | \, p,q,t)  &= - {\cal C}(1,2,\ldots,r) \times \frac{1}{\sigma_{p,q}\sigma_{q,t} \sigma_{t,p}} \sum_{i=1}^{r-1} \left( s_{iq} \frac{ \sigma_{q,p} \sigma_{i,r} }{\sigma_{i,q} \sigma_{r,p}} + (q\leftrightarrow t) \right)
\notag
\\
&\! \! \! \! \! \!  \! \! \! \! \! \!   \! \! \! \! \! \!  \! \! \! \! \! \!  =\sum_{i=1}^{r-1} (-1)^{r-i}   \! \! \! \! \! \!  \!    \sum_{\rho \in \{1,2,\ldots,i-1\}\atop{ \shuffle \{r-1,\ldots,i+1\}}} \! \! \! \! \! \!  \big[ s_{iq}  {\cal C}(\rho,i,q,t,p,r) - s_{it}  {\cal C}(\rho,i,t,q,p,r)  \big]  \ ,\label{mtrace8}
\end{align}
which translates to the amplitude relation (\ref{mtrace4}). The numerator factor $\sigma_{q,p}$ and its image under  $(q\leftrightarrow t)$ have been canceled against the three-particle Parke-Taylor factor $(\sigma_{p,q}\sigma_{q,t} \sigma_{t,p})^{-1}$.
\item For a cycle of length four, setting $(r{+}1,r{+}2,r{+}3,n) \rightarrow (p,q,t,u)$ yields
\begin{align}
{\cal J}^{\rm EYM}_{\{r,4\}}(1,2,\ldots,r \, | \, p,q,t,u)  &=  -{\cal C}(1,2,\ldots,r) {\cal C}(p,q,t,u) \sum_{i=1}^{r-1} \left( s_{iq} \frac{ \sigma_{q,p} \sigma_{i,r} }{\sigma_{i,q} \sigma_{r,p}} + (q\leftrightarrow t,u) \right) \notag \\
&\! \! \! \! \! \!  \! \! \! \! \! \!   \! \! \! \! \! \!  \! \! \! \! \! \!= \sum_{i=1}^{r-1} (-1)^{r-i}   \! \! \! \! \! \! \!    \sum_{\rho \in \{1,2,\ldots,i-1\} \atop{\shuffle \{r-1,\ldots,i+1\}}} \! \! \! \! \! \!  \big[ s_{iq}  {\cal C}(\rho,i,q,t,u,p,r) + s_{iu}  {\cal C}(\rho,i,u,t,q,p,r)  \label{mtrace9} \\
& \ \ \ \ \ \ \ \ \ \ \ -
 s_{it}  {\cal C}(\rho,i,t,q,u,p,r) -s_{it}  {\cal C}(\rho,i,t,u,q,p,r)  \big] \ ,
\notag
\end{align}
which translates to the amplitude relation (\ref{mtrace5}). The second term $\sim s_{i,t} \frac{ \sigma_{t,p} \sigma_{i,r} }{\sigma_{i,t} \sigma_{r,p}}$ in the first line requires the rearrangement 
${\cal C}(p,q,t,u)=-{\cal C}(p,t,q,u)-{\cal C}(p,t,u,q)$ of the four-particle Parke-Taylor factor to cancel the numerator $\sim \sigma_{t,p}$.
\item For two cycles of arbitrary length, we obtain
\begin{align}
{\cal J}^{\rm EYM}_{\{r,n-r\}}(1,2,\ldots,r \, | \, r{+}1,\ldots,n)  &=-  {\cal C}(1,2,\ldots,r)  {\cal C}(r{+}1,\ldots,n) \times \sum_{i=1}^{r-1} \sum_{j=r+2}^{n} s_{ij} \frac{ \sigma_{j, r+1} \sigma_{i,r} }{\sigma_{i,j} \sigma_{r,r+1}} \notag \\
&\! \! \! \! \! \!  \! \! \! \! \! \!   \! \! \! \! \! \!  \! \! \! \! \! \! 
\! \! \! \! \! \!  \! \! \! \! \! \!  
= \sum_{i=1}^{r-1} \sum_{j=r+2}^n (-1)^{j-i}  s_{ij}  \! \! \! \! \! \!  \sum_{\rho \in \{1,2,\ldots,i-1\} \atop{ \shuffle \{r-1,\ldots,i+1\}}}  \sum_{\tau \in \{j+1 ,\ldots,n\} \atop{\shuffle \{ j-1, \ldots, r+2 \}}}  \! \! \! \! \! \! \! \! {\cal C}(\rho,i,j,\tau,r{+}1,r)\ ,
\end{align}
which translates to the most general double-trace amplitude relation (\ref{mtrace2}). The Parke-Taylor factor ${\cal C}(r{+}1,\ldots,j,\ldots,n)$ has been expressed in a KK-basis of ${\cal C}(\ldots ,j,r{+}1)$ to cancel $ \sigma_{j, r+1}$ in the numerator.
\end{itemize} 

\subsection{An alternative representation}

Similar to the observations in section \ref{sec42}, KK-relations (\ref{shufflerel}) give rise to a variety of equivalent representations of double-trace amplitude relations. Repeating the above rewritings of (\ref{mtrace1}) in a frame where $\sigma_n \rightarrow \infty$ leads to the following alternative representation of (\ref{mtrace2}):
\begin{align}
{\cal A}_{\rm EYM}&(1,2,\ldots,r \, | \,r{+}1,\ldots,n) =   - \sum_{j=1}^{r-1} \sum_{\ell=r+2}^{n} (-1)^{n-\ell} s_{j\ell} \notag \\
& \times \!\!\!\!\!\! \sum_{\tau \in \{ r+2,\ldots, \ell-1 \} \atop { \shuffle \{n,n-1,\ldots,\ell+1 \} } } \sum_{\sigma \in \{1,2,\ldots, j-1\} \atop{ \shuffle \{r+1,\tau,\ell \} }} {\cal A}(\sigma,j,j{+}1,\ldots,r) \ .
\label{spcase}
\end{align}
Note that the sets $\tau$ from the first sum over shuffles enter the summation range of $\sigma$. The special cases
of (\ref{spcase}) with a small number of gluons in one of the cycles,
\begin{align}
{\cal A}_{\rm EYM}(1,2,\ldots,r \, | \,p,q) &=
- \sum_{j=1}^{r-1} s_{jq} \! \! \!  \sum_{ \sigma \in \{ p,q\} \atop { \shuffle \{ 1,2,\ldots,j-1 \} } } \! \! \!  {\cal A}(\sigma,j,j{+}1,\ldots,r) \\
{\cal A}_{\rm EYM}(1,2,\ldots,r \, | \,p,q,t) &=
 \sum_{j=1}^{r-1} s_{jq} \! \! \!  \sum_{ \sigma \in \{ p,t,q\} \atop { \shuffle \{ 1,2,\ldots,j-1 \} } } \! \! \!  {\cal A}(\sigma,j,j{+}1,\ldots,r) - (q\leftrightarrow t) \\
{\cal A}_{\rm EYM}(1,2,\ldots,r \, | \,p,q,t,u) &= \sum_{j=1}^{r-1} \Big\{ s_{jt} \! \! \!  \sum_{ \sigma \in \{ p,q,u,t\} \atop { \shuffle \{ 1,2,\ldots,j-1 \} } }  \! \! \! {\cal A}(\sigma,j,j{+}1,\ldots,r) \\
&\ \ \ \ \ \  - s_{jq} \! \! \!  \sum_{ \sigma \in \{ p,u,t,q \} \atop { \shuffle \{ 1,2,\ldots,j-1 \} } } \! \! \!  {\cal A}(\sigma,j,j{+}1,\ldots,r) \Big\} + (q\leftrightarrow u)  \ , \notag
\end{align}
are related to (\ref{mtrace3}) to (\ref{mtrace5}) by a sequence of KK relations.

\subsection{A double-trace counterpart of BCJ-relations}

While BCJ relations among single-trace amplitudes can be written in the form 
\cite{delaCruz:2015raa}
\beq
\sum_{l=1}^{n-1} (p \cdot x_l) {\cal A}(1,2,\ldots,l,p,l+1,\ldots,n) = 0 \ ,
\eeq
double-trace amplitudes satisfy a modified version of this relation,
\begin{align}
0&= \sum_{l=1}^{r-1} (p \cdot x_l) {\cal A}_{\rm EYM}(1,2,\ldots,l,p,l{+}1,\ldots,r \, | \, r{+}1,\ldots,n) \notag \\
&+\sum_{l=r+1}^{n-1} (p \cdot x_l) {\cal A}_{\rm EYM}(1,2,\ldots,r \, | \, r{+}1,\ldots,l,p,l{+}1,\ldots, n)  \label{BCJlike21}  \\
&- (p \cdot x_r) \,
\, \sum_{i=1}^{r-1} \sum_{j=r+2}^{n} (-)^{i-j} s_{ij} \sum_{\sigma  \in \{1,2,\ldots, i-1\} \atop{ \shuffle \{r-1,r-2,\ldots, i+1\} }}
\sum_{\tau \in \{  j-1,j-2,\ldots, r+2\} \atop{ \shuffle \{j+1,j+2,\ldots, n \}  }}
 {\cal A}(r, \sigma,i,j,\tau, r{+}1 ,p) \ ,
 \notag
\end{align}
with a single-trace admixture in the last line. For small numbers of particles, (\ref{BCJlike21}) specializes to
\begin{align}
0&= (p\cdot x_1)  {\cal A}_{\rm EYM}(1,p,2 \,|\, 3,4) +(p\cdot x_3)  {\cal A}_{\rm EYM}(1,2 \,|\, 3,p,4)+ 
(p\cdot x_2)  s_{14} {\cal A} (2,1,4,3,p)
\label{BCJlike22} \\
0&= (p\cdot x_1)  {\cal A}_{\rm EYM}(1,p,2,3 \,|\, 4,5) +(p\cdot x_2)  {\cal A}_{\rm EYM}(1,2,p,3 \,|\, 4,5) + (p\cdot x_4)  {\cal A}_{\rm EYM}(1,2,3 \,|\, 4,p,5) \notag \\
& 
+ (p\cdot x_3) \big[ s_{25} {\cal A} (3,1,2,5,4,p) -s_{15} {\cal A} (3,2,1,5,4,p) \big]
\label{BCJlike23} \\
0&= (p\cdot x_1)  {\cal A}_{\rm EYM}(1,p,2 \,|\, 3,4,5) +(p\cdot x_3)  {\cal A}_{\rm EYM}(1,2 \,|\, 3,p,4,5) + (p\cdot x_4)  {\cal A}_{\rm EYM}(1,2 \,|\, 3,4,p,5)\notag \\
& 
+ (p\cdot x_2) \big[ s_{14} {\cal A} (2,1,4,5,3,p) -s_{15} {\cal A} (2,1,5,4,3,p) \big] \ .
\label{BCJlike24}
\end{align}
Any instance of (\ref{BCJlike21}) can be verified by converting the double-trace amplitudes to single-trace expressions via (\ref{mtrace2}) and expanding the latter in a BCJ basis. In its general form, however, (\ref{BCJlike21}) remains conjectural at this point. All cases involving ${\cal A}(\ldots)$ of multiplicity $n\leq 7$ have been checked in generic dimensions, and we additionally performed numerical checks in four-dimensional MHV helicity configurations for up to nine points.


\section{Double-trace amplitude relations with one graviton}
\label{section:moremultitrace}

Following our discussion in the previous section, we shall now present the double-trace contributions to EYM amplitudes $ \mathcal{A}_{\rm EYM}(\{1,2,\ldots,r \, | \,r{+}1, \ldots,n\}, p) $ with one graviton labelled by $\{\ep_p,k_p\equiv p\} $. As will be derived in the remainder of this section, these mixed amplitudes boil down to their purely gluonic counterparts through the all-multiplicity formula
\begin{align}
&\mathcal{A}_{\rm EYM}(\{1,2,\ldots,r \, | \,r{+}1, \ldots,n\}, p)= \sum_{l=1}^{r-1} (\ep_p\cdot x_l)  
	 \, \mathcal{A}_{\rm EYM}(1,2,\ldots,l,p,l{+}1,\ldots,r\, | \,r{+}1, \ldots,n) \notag \\
& \ \ \ \ \ \ \ \ \ \ \ \ \ \ \ \ \ \ \ \  - \, (\ep_p\cdot x_r) \, \sum_{i=1}^{r-1} \sum_{j=r+2}^{n} (-)^{i-j} s_{ij} \sum_{\sigma  \in \{1,2,\ldots, i-1\} \atop{ \shuffle \{r-1,r-2,\ldots, i+1\} }} \sum_{\tau \in \{  j-1,j-2,\ldots, r+2\} \atop{ \shuffle \{j+1,j+2,\ldots, n \}  }}
 {\cal A}(r, \sigma,i,j,\tau, r{+}1 ,p)    \notag \\
 & \ \ \ \ \ \ \ \ \ \ \ \ \ \ \ \ \ \ \ \  +\sum_{l=r+1}^{n-1}   (\ep_p\cdot x_l)  
	\, \mathcal{A}_{\rm EYM}(1,2,\ldots,r\, | \,r{+}1, \ldots,l,p,l{+}1,\ldots,n) \ ,  \label{doubletraceonegravgeneral} 
\end{align}
with both single-trace and double-trace contributions on the right hand side. At low multiplicity, (\ref{doubletraceonegravgeneral}) specializes to
\begin{align}
\mathcal{A}_{\rm EYM}(\{1,2 \, | \,3,4\}, p) &= (\ep_p \cdot x_1) \, \mathcal{A}_{\rm EYM} ({1, p, 2}\,|\, {3, 4})   \label{dtex1} \\
& \! \! \!  \! \! \!  \! \! \!  \! \! \!  \! \! \!  \! \! \!  \! \! \!  \! \! \!  \! \! \!  \! \! \!  \! \! \!  \! \! \!  + 
(\ep_p \cdot x_3) \,\mathcal{A}_{\rm EYM}({1, 2} \,|\,  {3, p, 4})+ 
(\ep_p \cdot x_2) \, s_{14} \, \mathcal{A}({2, 1, 4, 3, p})\notag
\\
\mathcal{A}_{\rm EYM}(\{1,2,3 \, | \,4,5\}, p) &= (\ep_p \cdot x_1) \, \mathcal{A}_{\rm EYM}({1, p, 2, 3} \,|\, {4, 5}) + 
(\ep_p \cdot x_2) \,\mathcal{A}_{\rm EYM}({1, 2, p, 3} \,|\,  {4, 5})    \label{dtex2} \\
& \! \! \!  \! \! \!  \! \! \!  \! \! \!  \! \! \!  \! \! \!  \! \! \!  \! \! \!  \! \! \!  \! \! \!  \! \! \!  \! \! \! + 
(\ep_p \cdot x_4) \,\mathcal{A}_{\rm EYM}({1, 2, 3} \,|\,  {4, p, 5})  + 
 (\ep_p \cdot x_3) \, \big[ s_{25}  \,\mathcal{A}({3, 1, 2, 5, 4, p})  - s_{15} \, \mathcal{A}({3, 2, 1, 5, 4, p})     \big] \notag
\\
\mathcal{A}_{\rm EYM}(\{1,2 \, | \,3,4,5\}, p) &= (\ep_p \cdot x_1) \, \mathcal{A}_{\rm EYM}({1, p, 2} \,|\,  {3, 4, 5}) + 
(\ep_p \cdot x_3) \,\mathcal{A}_{\rm EYM}({1, 2} \,|\,  {3, p, 4, 5})  \label{dtex3}\\
& \! \! \!  \! \! \!  \! \! \!  \! \! \!  \! \! \!  \! \! \!  \! \! \!  \! \! \!  \! \! \!  \! \! \!  \! \! \!  \! \! \! + 
(\ep_p \cdot x_4) \,\mathcal{A}_{\rm EYM}({1, 2} \,|\,  {3, 4, p, 5})  + 
(\ep_p \cdot x_2) \, \big[ s_{14} \, \mathcal{A}({2, 1, 4, 5, 3, p})  - 
s_{15}\, \mathcal{A}({2, 1, 5, 4, 3, p})  \big] \ .\notag
\end{align}
Note that the zero-graviton double-trace EYM amplitudes in \eqref{doubletraceonegravgeneral} can be further reduced to a basis of single-trace amplitudes in YM by using the relation (\ref{mtrace2}) from the previous section. 

\subsection{The integrand}

The above formula (\ref{doubletraceonegravgeneral}) originates from the CHY integrand \cite{Cachazo:2014xea}
for EYM double-trace amplitudes with $\Tr_1\equiv \{1,2\ldots,r\} $ and $\Tr_2\equiv \{r{+}1,\ldots,n\} $:
\begin{align}
&\mathcal{J}^{\rm EYM}_{\{r,n-r\}+1}(\{1,2,\ldots,r \, | \,r{+}1, \ldots,n\}, p)= \cC(1,2,\ldots,r) \, \cC(r{+}1,\ldots,n) \notag \\
& \ \ \ \ \ \ \ \times \ \Biggl[ s_{12\ldots r}(\sum_{i=1}^n\frac{k_i\cdot \ep_p }{\sigma_{i,p}})  -(\sum_{i\in\Tr_1}\frac{k_i\cdot p}{\sigma_{i,p}})(\sum_{j\in\Tr_1}\frac{\sigma_j k_j\cdot \ep_p}{\sigma_{j,p}}) +(\sum_{i\in\Tr_1}\frac{k_i\cdot \ep_p}{\sigma_{p,i}})(\sum_{j\in\Tr_1}\frac{\sigma_j k_j\cdot p}{\sigma_{j,p}}) \Biggr]\nn\\
&= \cC(1,2,\ldots,r)\,  \cC(r{+}1,\ldots,n)  \, \Biggl[\sum_{i=1}^{r-1}\sum_{j=i+1}^{r} {\cal F}_{ji} \frac{\sigma_{i,j}}{\sigma_{i,p}\sigma_{p,j}}
+
s_{12\ldots r} \, C_{pp}\Biggr] \ .
\label{doubletraceonegravCHY}
\end{align}
In proceeding to the last line we have introduced a shorthand for the tensor structure
\beq
{\cal F}_{ij} \equiv (k_i\cdot p)(k_j\cdot \ep_p)- (k_i\cdot \ep_p)(k_j\cdot p) \ ,
\label{Fij}
\eeq
which is built from the linearized field-strength $p^{\mu} \ep_p^{\nu}-p^{\nu} \ep_p^{\mu}$ and therefore gauge invariant. 
In order to spell out the CHY integrand (\ref{doubletraceonegravCHY}), one of the traces has to be singled out in the general formula of \cite{Cachazo:2014xea}. That is why the symmetry $\{1,2,\ldots,r \} \leftrightarrow \{r{+}1, \ldots,n\}$ under exchange of the color traces is obscured in \eqref{doubletraceonegravgeneral}. Verifying this hidden exchange symmetry for explicit examples such as (\ref{dtex1}) to (\ref{dtex3}) serves as a stringent consistency check of our results.

Similar to the strategy in the previous sections, the goal is to incorporate the $\sigma$-dependence from the square bracket of (\ref{doubletraceonegravCHY}) into the Parke-Taylor factors $\cC(1,2,\ldots,r)$ and $\cC(r{+}1,\ldots,n)$. Repeating the techniques from earlier sections, one can easily arrive at
\begin{align}
	&\mathcal{J}^{\rm EYM}_{\{r,n-r\}+1}(\{1,2,\ldots,r \, | \,r{+}1, \ldots,n\}, p)= - \sum_{1=i<j}^{r} {\cal F}_{ij} \sum^{j-1}_{ l=i } \mathcal{C}(1, 2, \ldots, l, p, l{+}1, \ldots, r )
\,\mathcal{C}(r{+}1, \ldots,n)  \notag
\\
&  \ \ \ \ \ \ \ \ \ \ \ \  \ \ \ \ \ \  +
	s_{12\ldots r} \, \Bigg\{ \, \sum_{l=1}^{r-1} (\ep_p\cdot x_l)  
	 \, \mathcal{C}(1,2,\ldots,l,p,l{+}1,\ldots,r)\, \mathcal{C}(r{+}1, \ldots,n) \label{intDT} \\
& \ \ \ \ \ \ \ \ \ \ \ \ \ \ \ \ \ \ \ \ \ \ \  \ \ \ \ \ \  +\sum_{l=r+1}^{n-1}   (\ep_p\cdot x_l) 
	\, \mathcal{C}(1,2,\ldots,r) \, \mathcal{C}(r{+}1, \ldots,l,p,l{+}1,\ldots,n)  \notag \\
& \ \ \ \ \ \ \ \ \ \ \ \ \ \ \ \ \ \ \ \ \ \ \  \ \ \ \ \ \  + \, (\ep_p\cdot x_r) \, \frac{ \sigma_{r,r+1} }{\sigma_{r,p} \sigma_{p,r+1}}  \, \mathcal{C}(1,2,\ldots,r) \, \mathcal{C}(r{+}1, \ldots,n) \, \Bigg\}  \ .\notag
\end{align}
\begin{itemize}
\item The factors of $\frac{\sigma_{i,j}}{\sigma_{i,p}\sigma_{p,j}}$ along with ${\cal F}_{ji} $ only interact with the legs in $\cC(1,2,\ldots,r)$. KK relations can be applied to expand the latter in a basis of $\cC(\ldots,i,j)$ which paves the way for the insertion of the graviton leg and leads to the first line of (\ref{intDT}).
\item In the usual expansion of $C_{pp}$ for a single graviton,
\beq
C_{pp} = \sum_{i=1}^{r-1} (\ep_p\cdot x_i) \, \frac{ \sigma_{i,i+1} }{\sigma_{i,p} \sigma_{p,i+1}} + \sum_{i=r+1}^{n-1} (\ep_p\cdot x_i) \, \frac{ \sigma_{i,i+1} }{\sigma_{i,p} \sigma_{p,i+1}} + (\ep_p\cdot x_r)\, \frac{ \sigma_{r,r+1} }{\sigma_{r,p} \sigma_{p,r+1}}\, ,
\label{C00double}
\eeq
all of the terms except for the last one conspire with one of $\cC(1,2,\ldots,r)$ or $\cC(r{+}1,\ldots,n)$ to yield an insertion of the graviton leg. This builds up the second and third line of (\ref{intDT}).
\item The last term in (\ref{C00double}) requires special attention since $\cC(1,2,\ldots,r) \frac{ \sigma_{r,r+1} }{\sigma_{r,p} \sigma_{p,r+1}} \cC(r{+}1,\ldots,n)$ does not relate to products of Parke-Taylor factors in an obvious manner. In the next section, we fill fix its net contribution to the amplitude relation (\ref{doubletraceonegravgeneral}) indirectly by imposing gauge invariance. 
\end{itemize}

\subsection{From the integrand to amplitudes}
\label{sec82}

In contrast to the simple conversion rule $\cC(1,2,\ldots,n) \rightarrow {\cal A}(1,2,\ldots,n)$ for single-trace amplitudes under the CHY measure, the products of Parke-Taylor factors in (\ref{intDT}) require an additional Mandelstam variable
$s_{12\ldots r} \,\cC(1,2,\ldots,r)  \cC(r{+}1,\ldots,n) \rightarrow {\cal A}_{\rm EYM}(1,2,\ldots,r \, | \, r{+}1,\ldots,n)$ 
to yield double-trace amplitudes via (\ref{mtrace1}). In the third line of (\ref{intDT}), the prefactor of $s_{12\ldots r}$ is manifestly compatible with the partition of legs among the Parke-Taylor factors, but the first two lines require a more careful analysis. Leaving the overall $\mathcal{C}(r{+}1, \ldots,n)$ aside, we have
\begin{align}
& - \sum_{1=i<j}^{r} {\cal F}_{ij} \, \sum^{j-1}_{ l=i } \mathcal{C}(1, 2, \ldots,  l, p, l{+}1, \ldots, r )\label{congkao} \\
&\ \ \  = \sum_{l=1}^{r-1} \left[ (p \cdot x_r) (\ep_p\cdot x_l) - (\ep_p \cdot x_r) (p\cdot x_l) \right]
\, \mathcal{C}(1,2,\ldots,l,p,l{+}1,\ldots,r) \ , \notag
\end{align}
after rewriting (\ref{Fij}) in terms of $(\ep_p \cdot x_j)$. This can be straightforwardly proven by considering 
each Parke-Taylor factor in the sum, for instance $ \mathcal{C}(1,\ldots,l,p,l{+}1,\ldots,r)$, and checking that
the $\ep_p$-dependent coefficient of this particular Parke-Taylor factor is identical on both sides of the equation. 
We can take advantage of (\ref{congkao}) to rearrange the first two lines of (\ref{intDT}):
\begin{align}
&- \sum_{1=i<j}^{r} {\cal F}_{ij} \, \sum^{j-1}_{ l=i } \mathcal{C}(1, 2, \ldots,  l, p, l{+}1, \ldots, r ) + s_{12\ldots r} \,  \sum_{l=1}^{r-1} (\ep_p\cdot x_l)  
	 \, \mathcal{C}(1,\ldots,l,p,l{+}1,\ldots,r) \label{partDT} \\
&\ \ \ = s_{p,12\ldots r} \,  \sum_{l=1}^{r-1} (\ep_p\cdot x_l)  
	 \, \mathcal{C}(1,\ldots,l,p,l{+}1,\ldots,r) - (\ep_p \cdot x_r) \sum_{l=1}^{r-1}  \sum_{j=1}^l \, s_{pj}\, \mathcal{C}(1,\ldots,l,p,l{+}1,\ldots,r) \ . \notag
\end{align}
With (\ref{partDT}) and the third line of (\ref{intDT}), we arrive at 
the following partial answer for the desired amplitude relation
\begin{align}
&\mathcal{A}_{\rm EYM}(\{1,2,\ldots,r \, | \,r+1, \ldots,n\}, p)= \sum_{l=1}^{r-1} (\ep_p\cdot x_l)  
	 \, \mathcal{A}_{\rm EYM}(1,\ldots,l,p,l{+}1,\ldots,r\, | \,r{+}1, \ldots,n) \notag \\
& \ \ \ \ \ \ \ \ \ \ \ \ \ \ +\sum_{l=r+1}^{n-1}   (\ep_p\cdot x_l)  
	\, \mathcal{A}_{\rm EYM}(1,\ldots,r\, | \,r{+}1, \ldots,l,p,l{+}1,\ldots,n)  + (\ep_p \cdot x_r) (\ldots) \ ,  \label{partialDT} 
\end{align}
where the unevaluated coefficient of $(\ep_p \cdot x_r)$ stems from the fourth line of (\ref{intDT}) and the last term of (\ref{partDT}). Although the associated $\sigma$-dependences can be similarly rearranged via scattering equations, here we proceed with an alternative method by imposing gauge invariance under $\epsilon_p \rightarrow p$. 
Hence, the ellipsis in (\ref{partialDT}) along with $(\ep_p \cdot x_r)$ can be inferred as the unique gauge invariant completion
\begin{align}
	&\mathcal{A}_{\rm EYM}(\{1,2,\ldots,r \, | \,r{+}1, \ldots,n\}, p) \, \big|_{ (\ep_p\cdot x_r) }= - \frac{1}{(p \cdot x_r)} \notag \\
	& \ \ \ \times \
	\Biggl\{ \, \sum_{l=1}^{r-1}  (p \cdot x_l) 
	 \, \mathcal{A}_{\rm EYM}(1,2,\ldots,l,p,l{+}1,\ldots,r\, | \,r{+}1, \ldots,n) \label{gaugeinv} \\
	 & \ \ \ \ \ \ \ 
	  +\sum_{l=r+1}^{n-1}   (p \cdot x_l)	
	  \, \mathcal{A}_{\rm EYM}(1,2,\ldots,r\, | \,r{+}1, \ldots,l,p,l{+}1,\ldots,n)   \, \Biggr\} 
		\notag \\
&= - \sum_{i=1}^{r-1} \sum_{j=r+2}^{n} (-)^{i-j} s_{ij} \sum_{\sigma  \in \{1,2,\ldots, i-1\} \atop{ \shuffle \{r-1,\ldots, i+1\} }}
\sum_{\tau \in \{  j-1,\ldots, r+2\} \atop{ \shuffle \{j+1,\ldots, n-1,n \}  }}
 {\cal A}(r, \sigma,i,j,\tau, r{+}1 ,p)  \ . \notag 
\end{align}
In proceeding to the last line, we have used the relation (\ref{BCJlike21}) between gluon amplitudes of single- and double-trace type to cancel the spurious pole as $(p\cdot x_r) \rightarrow 0$. Hence, (\ref{partialDT}) and (\ref{gaugeinv}) complete the derivation of the general amplitude relation (\ref{doubletraceonegravgeneral}).

\section{Conclusion and outlook}

In this work, we have presented new relations to reduce EYM amplitudes involving up to three gravitons and up to two color traces to partial amplitudes of pure YM theories. From their derivation in the dimension-agnostic CHY formalism, the results are valid for external bosons in any number of spacetime dimensions. Moreover, the CHY formula for gluonic YM amplitudes from a Pfaffian is supersymmetrized\footnote{This has been established in \cite{Gomez:2013wza} by comparing the vertex operators and their operator product expansions in the pure-spinor incarnation of the CHY formalism \cite{Berkovits:2013xba} and the open pure-spinor superstring \cite{Berkovits:2000fe}.} by the open-string correlators in pure-spinor superspace \cite{Mafra:2011kj, Mafra:2011nv}. Hence, our results extend to any superamplitude which descends from ten-dimensional SYM coupled to half-maximal supergravity.

This work raises a variety of follow-up questions: Most obviously, it would be desirable to extend the amplitude relations to any number of color traces and graviton states potentially
uncovering a recursive structure. Another interesting direction is to consider the generalization of such relations at loop level. It is actually easy to see that the simple identity (\ref{master}) should be violated at loop level for instance by considering the simplest one-loop amplitude in four dimensions: The four-point all-plus helicity amplitude. For this particular case, we see that the right hand side of (\ref{master}) is in fact not gauge invariant. This follows either from the explicit form of four-point all-plus helicity gluon amplitude at one loop \cite{Bern:1995db} or from the known fact that the amplitudes generated from the higher-dimensional term $F^4$ \cite{Mafra:2012kh, Broedel:2012rc} do not obey the BCJ relations. One can draw the same conclusion by considering the IR divergent part of loop amplitudes. Finally, it would be interesting to re-derive the results of this work from the heterotic string and compare the additional string corrections with the open-string results of \cite{Stieberger:2009hq, Stieberger:2015vya, Stieberger:2016lng}. One might speculate about new connections between the tree amplitudes among gluons and gravitons in different string theories along the lines of \cite{Stieberger:2014hba, Huang:2016tag}.

\subsubsection*{Acknowledgments}

We wish to thank Marco Chiodaroli for helpful comments on the draft and
Johannes Br\"odel for sharing several instrumental Mathematica tools. 
Moreover, we are grateful to Stephan Stieberger, Tomasz Taylor and Wadim Wormsbecher for discussions.
CW thanks Humboldt University for hospitality.

\appendix
\section{Further details on the four-graviton case}
\label{app4G}
In this appendix, we provide some more details for the special case of four gravitons in section \ref{sec6}. Specifically we present the key steps to express the most nontrivial term in \eqref{eq:fourgravitonlist} along with $ (\ep_p\cdot\ep_q)(\ep_r\cdot\ep_t)$ in terms of $(n{+}4)$-particle Parke-Taylor factors. 

For this tensor structure, the accompanying dependence on the $\sigma$-variables takes the form, 
\be
\mcC(1,2,\ldots,n)\frac{s_{pq}}{\sigma_{p,q}^2}\frac{s_{rt}}{\sigma_{r,t}^2}=\mcC(1,2,\ldots,n)\frac{s_{pq}}{\sigma_{p,q}^2}\frac{1}{\sigma_{r,t}}\biggl(\underbrace{\frac{s_{tp}}{\sigma_{t,p}}}_{A_1}+\underbrace{\frac{s_{tq}}{\sigma_{t,q}}}_{A_2}+\underbrace{\sum_{i=1}^{n}\frac{s_{ti}}{\sigma_{t,i}}}_{A_3}\biggr),
\label{A123}
\ee
where the sum of terms $ \{A_1,A_2,A_3\} $ inside the parenthesis results from using the scattering equation with respect to the particle labeled $t$ to rearrange the term $ \frac{s_{rt}}{\sigma_{r,t}} $. We note that $ A_1 $
 is related to $ A_2 $ via $ p\leftrightarrow q $. So we just discuss the simplification of the last two terms below.
 \paragraph*{Second term $ A_2 $:} As in our previous discussions we will choose a frame $ \sigma_n\rightarrow\infty $, for which the Parke-Taylor factor in (\ref{A123}) is denoted as $ \cC(1,2,\ldots,\hat{n}) $. Using this frame and the scattering equation for particle $ p $ in $\frac{s_{pq}}{\sigma_{p,q}^2}  $,  the $ A_2 $-contribution to (\ref{A123}) can be written as,   
 \be
  \mcC(1,2,\ldots,n)\frac{s_{pq}}{\sigma_{p,q}^2}\frac{s_{rt}}{\sigma_{r,t}^2} \, \Big|_{A_2} \rightarrow
 \frac{s_{tq}}{\sigma_{r,t}\sigma_{t,q}\sigma_{p,q}}\biggl( \underbrace{ \frac{s_{rp}}{\sigma_{r,p}}}_{B_1}+ \underbrace{\frac{s_{tp}}{\sigma_{t,p}}}_{B_2}+\underbrace{ \sum_{j=1}^{n-1}\frac{s_{jp}}{\sigma_{j,p}}  }_{B_3}\biggr)\cC(1,2,\ldots,\hat{n}) \ , 
 \ee
where the first term $B_1$ can be identified as $ s_{tq}\,s_{rp}\,\cC(r,t,p,q) \,\cC(1,2,\ldots,n) $ and rewritten in terms of $(n{+}4)$-particle Parke-Taylor factors via (\ref{mtrace9}).  Moreover, by repeated use of the partial-fraction identity $ \frac{1}{\sigma_{a,b}\sigma_{b,c}}=\frac{1}{\sigma_{a,c}\sigma_{b,c}}+\frac{1}{\sigma_{a,b}\sigma_{a,c}} $, the third term $B_3$ can also be simplified to
 \be
   \mcC(1,2,\ldots,n)\frac{s_{pq}}{\sigma_{p,q}^2}\frac{s_{rt}}{\sigma_{r,t}^2} \, \Big|_{A_2,B_3}=
s_{tq} \sum_{j=1}^{n-1}s_{jp} \sum_{\sigma \in \{r,t,q,p\} \atop{\shuffle \{ 1,2,\ldots, j-1\}} } {\cal C}(\sigma,j,j{+}1,\ldots,n) \ .
 \ee
Finally, the second term $B_2$ can be addressed using the identity
 \begin{align}
 &s_{pqrt}\cC(1,2,\ldots,n)\frac{\sigma_{n,t}}{\sigma_{n,r}\sigma_{r,t}}\cC(t,p,q)= \sum_{l=1}^{n-1}(x_l\cdot r) \,\cC(1,2,\ldots,l,r,l{+}1,\ldots,n) \, \cC(t,p,q)\\
&\ \ \ \ \ \ \ \ \ \ \ \ \ \ \ \ \ \ \ \ \ + \sum_{i=1}^{n-1}(-1)^{i-n}\sum_{\sigma \in \{1,2,\dots,i-1\} \atop{\shuffle \{ n-1,\ldots,i+1\}} }\biggl(s_{iq}{\cal C}(n,\sigma,i,q,p,t,r) - s_{ip}{\cal C}(n,\sigma,i,p,q,t,r)  \biggr)\ , \nn
 \end{align}
which is implicit in the analysis of section \ref{sec82}. The second line is already of the desired Parke-Taylor type, and the products in the first line can be brought into the same form via (\ref{mtrace8}).

 \paragraph{Last term $ A_3 $:} In a frame where $\sigma_n\rightarrow \infty$, the techniques of section \ref{sec43} can be applied to the last term $A_3$ in (\ref{A123}),
 \be
 \mcC(1,2,\ldots,n)\frac{s_{pq}}{\sigma_{p,q}^2}\frac{s_{rt}}{\sigma_{r,t}^2} \, \Big|_{A_3}= 
 \frac{s_{pq}}{\sigma_{p,q}^2}\sum_{i=1}^{n-1}s_{ti}\sum_{\sigma \in \{r,t\} \atop{\shuffle \{ 1,2,\ldots,i-1\}}} {\cal C}(\sigma,i,i{+}1,\ldots,n) \, .
  \ee
Any term on the right hand side is a product of the form ${\cal C}(p,q){\cal C}(\ldots)$ and can then be written 
in terms of single Parke-Taylor factor of length $n{+}4$ via (\ref{mtrace7}).
These are all the terms needed to simplify the contribution $\sim (\ep_p\cdot\ep_q)(\ep_r\cdot\ep_t)$ to the four-graviton case.

\bibliographystyle{nb}
\bibliography{soft}

\end{document}